\RequirePackage{fix-cm}
\documentclass[11pt,a4paper]{article}
\usepackage{graphicx}
\usepackage[reqno]{amsmath}
\usepackage{amssymb}
\usepackage{mathtools}
\usepackage{mathrsfs}

\usepackage[hyphens]{url}
\usepackage{hyperref}

\usepackage{graphicx}
\usepackage{subfig}

\usepackage{multirow}

\usepackage{authblk}

\providecommand{\keywords}[1]
{
  \small	
  \textbf{\textit{Keywords--}} #1
}

\providecommand{\acknowledgements}[1]
{
  \small	
  \textbf{Acknowledgements} #1
}

\title{A multi-task learning-based optimization approach for finding diverse sets of material microstructures with desired properties and its application to texture optimization}
\author[1]{Tarek Iraki}
\author[2]{Lukas Morand}
\author[3]{Johannes Dornheim}
\author[1]{Norbert Link}
\author[2]{Dirk Helm}
\affil[1]{Intelligent Systems Research Group ISRG, Karlsruhe University of Applied Sciences, Karlsruhe, Germany (e-mail: {tarek.iraki, norbert.link}@h-ka.de)}
\affil[2]{Fraunhofer Institute for Mechanics of Materials IWM, Freiburg, Germany (e-mail: {lukas.morand, dirk.helm}@iwm.fraunhofer.de)}
\affil[3]{Institute for Applied Mechanics - Computational Materials Sciences IAM-CMS, Karlsruhe Institute of Technology, Karlsruhe, Germany (email: johannes.dornheim@kit.edu)}

\begin{document}

\date{}
\maketitle

\begin{abstract}
The optimization along the chain processing-structure-properties-performance is one of the core objectives in data-driven materials science. In this sense, processes are supposed to manufacture workpieces with targeted material microstructures. These microstructures are defined by the material properties of interest and identifying them is a question of materials design. In the present paper, we addresse this issue and introduce a generic multi-task learning-based optimization approach. The approach enables the identification of sets of highly diverse microstructures for given desired properties and corresponding tolerances. Basically, the approach consists of an optimization algorithm that interacts with a machine learning model that combines multi-task learning with siamese neural networks. The resulting model (1) relates microstructures and properties, (2) estimates the likelihood of a microstructure of being producible, and (3) performs a distance preserving microstructure feature extraction in order to generate a lower dimensional latent feature space to enable efficient optimization. The proposed approach is applied on a crystallographic texture optimization problem for rolled steel sheets given desired properties.


%\keywords{Machine Learning \and Distance preserving feature extraction \and Multi task learning \and Siamese neural networks \and Evolutionary algorithm \and Rolling texture \and Crystal plasticity}

\end{abstract}

\section{Introduction}
\label{intro}

%TODO: variablennamen abgleichen: $\alpha$, $\beta$, feature-dimensionen $N$, $M$, Material-Parameters $P$, Population $P$, anzahl samples in dataset $K$, NN-parameter $\theta$,  ...

\subsection{Motivation}

The demand for more and more specific and individually designed products with certain performance requirements has become a driving force in the world of manufacturing. For this reason, the optimization along the causal chain processing-structure-properties-performance \cite{olson1997computational} became a fast growing research topic in the field of integrated computational materials engineering (ICME) \cite{panchal2013key}. Nowadays, such optimization problems can be solved efficiently with the help of machine learning techniques \cite{ramprasad2017machine}. On this background, in a previous work, we investigated the use of reinforcement learning for finding optimal processing routes in a simulated metal forming process aiming to produce microstructures with targeted crystallographic textures \cite{dornheim2020structureguided}. To bridge the remaining gap between microstructures and desired properties, we focus in this work on solving materials design problems. These are to identify appropriate material microstructures or microstructural features (e.g. the crystallographic texture) for given desired properties. It is thereby of particular importance to identify sets of near-optimal and preferably diverse microstructures in order to guarantee a robust design \cite{mcdowell2007simulation}.

\subsection{Paper structure}
In the following we summarize the related work and point out the contribution of this paper. In Section \ref{sec:methods}, first, we describe the siamese multi-task learning and optimization approach. Then, we introduce the fundamentals in materials modeling that are needed for the purpose of this work. After that, in Section \ref{sec:results}, the results are shown when applying the approach to a texture optimization problem for steel sheets. In Section \ref{sec:discussion}, the presented results are discussed. Finally, in Section \ref{sec:conclusion}, we summarize our findings and give an outlook on further research.

\subsection{Related work}

A recent and very generic approach to solve materials design problems is the microstructure sensitive design (MSD) approach introduced in \cite{adams2001microstructure}. Following \cite{fullwood2010microstructure}, MSD can be described by the seven steps. First, the properties of interest as well as candidate materials have to be defined. After that, a suitable microstructure definition is applied for these materials yielding a microstructure design space. On this basis, relevant homogenization relations are identified and applied over the whole design space. The resulting properties closure can be used to select desired properties, which are then mapped back to the microstructure design space in order to identify optimal microstructures. The last step of MSD aims to determine processes and processing routes needed to produce the identified microstructure. 
%\textbf{TI NEU: The useage of precalculated processing routes leads to a rigid coupling between process routes and properties. This does not make flexible process control possible.}

%\fullwood nur vorgegebene vorberechnete prozesspfade, starre kopplung prozesspfade und eigenschaften. graphen traversieren. prozesspfade schon festgekoppelt mit vorg. eigenschaften und damit flexible prozess-regelung nicht möglich.}

The works by Adams et al. \cite{adams2001microstructure} and Kalidindi et al. \cite{kalidindi2004microstructure} instantiate the MSD approach for texture optimization. The first one describes how optimal crystallographic textures can be identified in order to improve the deformation behavior of a compliant beam. In the latter, a similar approach is shown to optimize the crystallographic texture for the design of an orthotropic plate. The core of both approaches lies in the usage of a lower dimensional spectral representation of the orientation distribution, cf. \cite{bunge2013texture}. For more complex microstructure representations, like two-point correlations, feature extraction methods can be applied to reduce the dimensionality. Methods that are for example used in the context of materials design are principal component analysis (PCA) \cite{paulson2017reduced, gupta2015structure} and multidimensional scaling \cite{JUNG2019-MDS}. A general review of dimensionality reduction techniques can be found in
\cite{van2009dimensionality}.

Besides the MSD approach, also machine learning-based approaches for crystallographic texture optimization exist. \cite{liu2015predictive} and \cite{paul2019microstructure} describe iterative sampling approaches that interact with crystal plasticity simulations aiming to identify crystallographic textures for given desired properties. Therefore, an initial set of texture-properties tuples is generated. Via supervised learning, significant features of the parameterized orientation distribution (and in \cite{liu2015predictive} also regions) are identified that yield optimal or near-optimal solutions. Based on the identified features and regions, new texture-properties data points are sampled in order to get closer to the optima.

%Furthermore, a fundamental characteristic of our approach is that it relies on distance measures between microstructures which can be preserved in the lower dimensional design space. This enables us to force optimization algorithms to diversify the set of identified microstructures. %LM: hab ich so in der art in die motivation geschrieben 

%The approach presented in \cite{kuroda2004texture} uses a real-coded genetic algorithm \cite{goldberg1990real} that interacts with a crystal plasticity model in order to find optimal combinations of typical fcc rolling texture components (Cu, Brass, S, Cube and Goss) for given target properties. The algorithm used therein starts with an initial set of textures with different fractions of these components. The set of textures evolves iteratively by combining them using operators such as mutation, crossover and selection \cite{herrera1998tackling}. 

Another approach for identifying optimal textures is described in \cite{kuroda2004texture}. Therein, a real-coded genetic algorithm \cite{goldberg1990real} is described that interacts with a crystal plasticity model in order to find optimal combinations of typical fcc rolling texture components (Cu, Brass, S, Cube and Goss) for given desired properties. The algorithm starts with an initial set of textures consisting of different fractions of these components. The set of textures evolves iteratively by combining them using operators such as mutation, crossover and selection \cite{herrera1998tackling}.

Summarized, for the solution of microstructure design problems, a linkage from properties to microstructures is required. Such a linkage is often achieved by genetic or optimization algorithms that interact with numerical simulations. However, as these algorithms generally need a lot of function evaluations, it is not reasonable to apply them to complex numerical simulations directly. Instead, the performance can be increased by using numerically simpler surrogate models \cite{simpson2001metamodels}. Typically, these are supervised learning models that learn the input-output relations of the numerical simulation under consideration.

To run optimization algorithms in combination with supervised learning models it is necessary to limit the region in which they operate to the region, which is covered by the training data. One way to achieve this is by training unsupervised learning methods on the input data, as it is for example done in \cite{Jung2019-ocsvm} using support vector machines (SVM). From a machine learning perspective such an approach can be seen as anomaly detection. Anomaly detection aims to separate data that is characteristically different from the known data of the sample data set, which has been used for training. An extensive overview of anomaly detection methods is given in \cite{Chandola2009-AnomalyDetection}. Moreover, \cite{Deep-Learning-for-Anomaly-Detection-Survey-2019} gives an overview on recent deep learning-based approaches for anomaly detection, from which we want to point out neural network-based autoencoders \cite{HintonSalakhutdinov2006b}, which fit especially well into multi-task learning (MTL) schemes other than SVMs.

Autoencoder approaches assume that features of a data set can be mapped into a lower dimensional latent feature space, in which the known data points differ substantially from unknown data points. By backmapping into the original space, anomalies can be identified by evaluating the reconstruction error, see for example \cite{Anomaly-Detection-Using-Autoencoders-Sakurada-2014}. In \cite{Anomaly-Detection-Using-Autoencoders-Sakurada-2014} it is also shown that autoencoder networks are able to detect subtle anomalies, which cannot be detected by linear methods like PCA. Furthermore, autoencoder networks require less complex computations compared to a nonlinear kernel-based PCA. 

\subsection{Contribution}
\label{sec:Contribution}
In the present paper, we introduce a generic MTL-based optimization approach to efficiently identify sets of microstructures, which are highly divers and producible by a process. The approach is based on an optimization algorithm interacting with a machine learning model that combines MTL \cite{caruana1997mtl} with siamese neural networks \cite{bromley1993siamese}. In contrast to \cite{liu2015predictive,paul2019microstructure} and also to \cite{kuroda2004texture}, in our approach a surrogate model is set up in order to replace the numerical simulation, which maps microstructures to properties. The microstructure-properties mapping can be executed efficiently by means of the surrogate model within the optimization procedure. 

%For the optimization, it is important that identified microstructures are producible. 
To address the issue of a producibility, we include a neural network in the MTL structure, which estimates the validity of a microstructure in the sense of being producible. The efficiency of the optimization is further increased by transforming the microstructure representation into a lower dimensional latent feature space, which is formed by a non-linear data-driven autoencoder.
%The lower dimensional feature space is formed by a non-linear data-driven autoencoder, what enables the optimizer to find appropriate microstructure representations efficiently. Afterwards, the identified latent features can be decoded to corresponding microstructures by the decoder-part of the autoencoder. 
The resulting lower dimensional latent feature space delivers the input for the three neural networks: The first network maps the features to properties (surrogate model), the second network estimates the validity and the third network is the decoder-part of the autoencoder. 
%The simultaneous multi-task learning of the encoder and the attached three multi-task networks ensures that the lower dimensional latent feature space is optimal for all tasks.
As learning takes place simultaneously for the encoder and the attached tasks, it is ensured that the lower dimensional feature space is optimal for all tasks. In addition, we enforce the latent feature space to preserve microstructure distances by employing a siamese neural network and multidimensional scaling. On this basis, we force the optimizer to find a diverse set of solutions in the latent feature space.

%the optimization in the latent feature space aims to yield a set of highly diverse microstructures. We achieve this goal by enforcing the latent feature space to preserve the distances of the microstructure space by employing a siamese neural network and multidimensional scaling, and by forcing the optimizer to find a diverse set of solutions in the latent feature space. %Our approach is instantiated and validated at a texture optimization problem for rolled steel sheets. 

%The optimization problem is inspired by the work of Kuroda and Ikawa \cite{kuroda2004texture}.
%\textbf{todo tarek: hier einache aner bessere formulierung für was MTL gut ist hinsichtlich des lower dimensional design spaces; bei der bildung der features wird nur die varianz in den daten berücksichtigt. aber nicht die beziehung der eigenschaften. pca möglichst viel varianz der daten darstellen. kann im gegensatz stehen zur beziehung zu den eigenschaften. bei der bildung der features spielt die beziehung zu den eigenschaften keine rolle. zwar niedrigdim mekrmalsraum aber die gewonnen merkmalen keine beziehung zu den featurs.}

% KOMMENTAR: das war NL zu detailliert, zu nah an der loesung. deswegen raus, koennen wir an anderen stellen noch  wiederverwenden. bspw. contribution?

%, in which an approach based on genetic algorithms is presented that is used to identify optimal crystallographic textures for given desired properties of rolled aluminum sheets.

%on the validity of input microstructures in terms of the underlying data. 

%\input{Related-Work}

\section{Methods}
\label{sec:methods}
% Ab ende der methoden nur noch texture

%\textbf{TODO: dimensionen checken. rausschreiben} 

\subsection{Materials design via siamese multi-task learning (SMTL) and optimization}
\label{sec:concept}

\subsubsection{General Concept}
\label{sec:general_concept}

\begin{figure}[h]
	\centering
  	\includegraphics[width=0.95\linewidth]{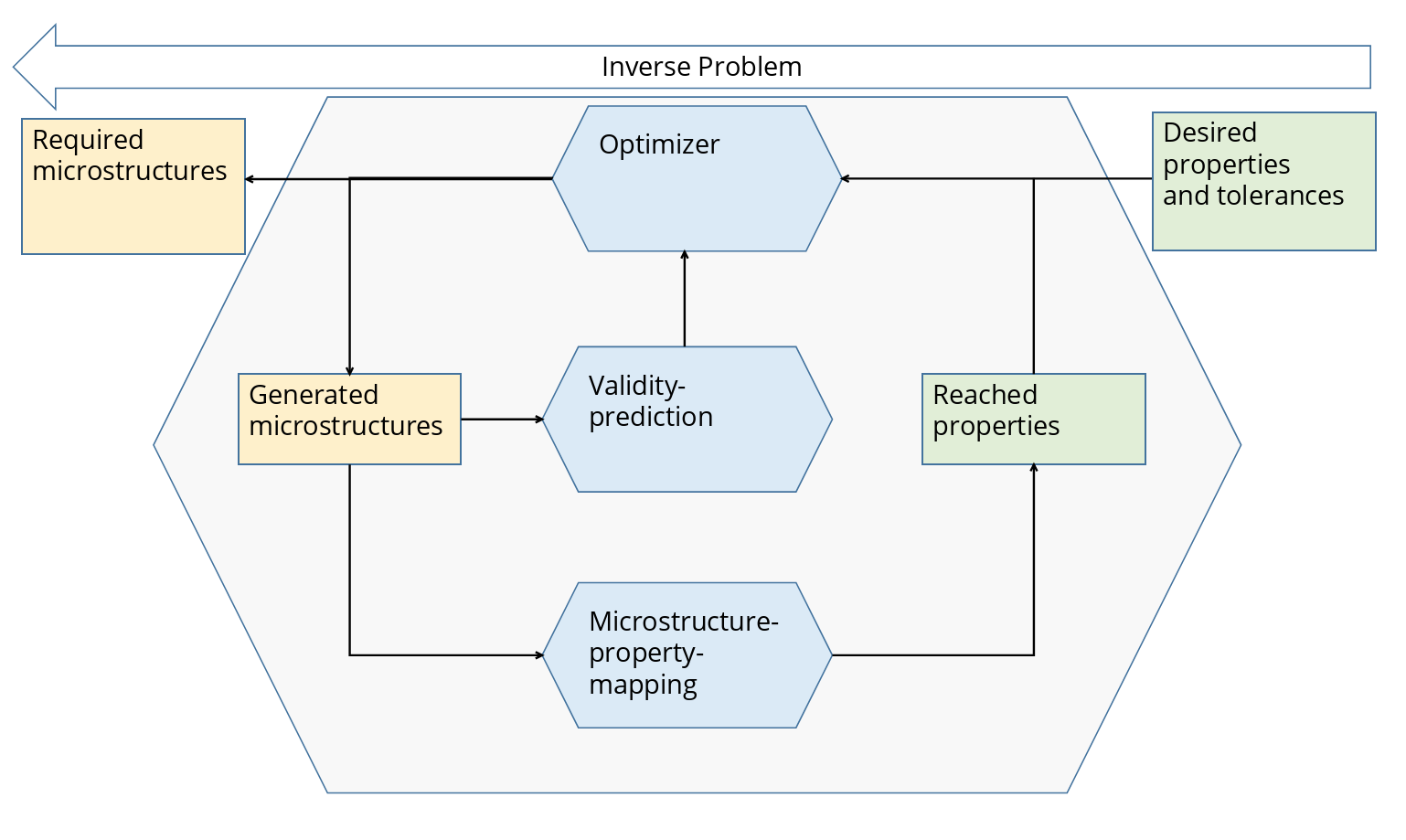}
	\caption{General concept of the MTL-based optimization approach to solve materials design problems.}
	\label{img:Solution-components}
\end{figure}

First of all, we present the general concept of our MTL-based optimization approach. The approach can be applied to general materials design problems and starts by defining the desired properties and corresponding tolerances. This defines a target region, for which the approach is supposed to identify a diverse set of microstructures. The approach is schematically depicted in Fig. \ref{img:Solution-components} and basically consists of three components: optimizer, microstructure-properties mapping (\textit{m-p-m}) and validity-prediction (\textit{v-p}). The optimizer generates candidate microstructures that minimize the combined costs, which result from evaluations based on the \textit{m-p-m} and \textit{v-p} components.

The \textit{m-p-m} component assigns properties to a candidate microstructure. The deviation of the assigned properties to the target region determines the cost. In general, the \textit{m-p-m} component can be realized by a numerical simulation. However, since numerical simulations are computationally expensive, a surrogate model is used instead. The surrogate model is realized by a regression model that learns the relations from a priori generated microstructure-properties data. 

\begin{figure}
	\centering
  	\includegraphics[width=0.95\linewidth]{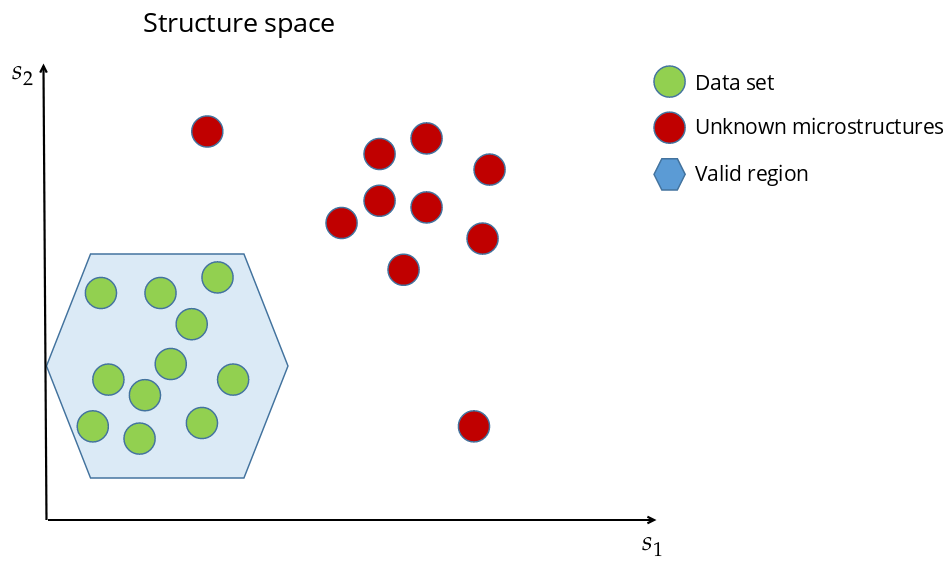}
	\caption{Schematic illustration of a set of generated microstructures in a structure space $s_1, s_2$. The valid region is the part of structure space, which is supported by the sampled data. 'Unknown' microstructures are located outside the valid region.}
	\label{img:Reachable-region}
\end{figure}

The \textit{v-p} component is realized by an anomaly detection method which determines the validity of a candidate microstructure by comparing it to the set of valid microstructures. The \textit{v-p} component returns a value that can be seen as an estimate of a candidate microstructure being an element of the microstructure set under consideration. This is for example the set, which can be produced by a dedicated process (e.g. rolling). %We call this the set of \textit{valid} microstructures. 
The value returned by the \textit{v-p} component defines the validity cost and drives the optimizer solution to a valid microstructure region, which is illustrated in Fig. \ref{img:Reachable-region}. Besides, such a microstructure region can also be identified by a further optimization algorithm that interacts with a numerical simulation of the dedicated process, however, such an approach suffers from high computational costs.
%\textbf{TODO lukas: vllt koennen wir hier doch noch die idee des zweiten surrogaten modells aufgreifen, da es zum generellen modell passt. wi bspw. bei den prozess daten? als modell gerlernt um aufwendige simulationen und optimierungen zu erspraren.}

%which represents an un-likelihood \textbf{anderes wort, bessere formulierung. estimate? distance measure?} of a microstructure being an element of the microstructure set under consideration.

%Both cost functions are set to zero, when they fall below a threshold value \textbf{trifft nur für vp zu. sonst ist es eine region}. This makes the optimizer contract the solutions in valid regions of the properties space and the microstructure space respectively. 

The two components \textit{m-p-m} and \textit{v-p} can be realized by training two separate machine learning models. However, when the training procedures are isolated from each other, the models are not able to mutually access information already learned by the other model. Therefore, we combine the two components as tasks into one MTL model \cite{caruana1997mtl}. Both tasks have a common backbone (the feature extraction part of a network) and different heads (feature processing part of a network) operating on the backbone output. The backbone output vectors form the so-called latent feature space. The proposed MTL approach furthermore uses the backbone as an encoder network of an autoencoder, where the decoder is also attached to the latent feature space with the purpose to reconstruct the input pattern of the backbone. This is achieved by adding the reconstruction of the microstructures from the latent feature space as a third task. In the MTL approach, all three tasks are represented by a single neural network-based model. The weights of the model are trained simultaneously based on a combined loss function. After training the MTL model, the optimizer can operate very efficiently in the lower dimensional latent feature space. The remainder of this section presents the optimization approach and the MTL approach in detail, as well as an extension based on siamese neural networks \cite{bromley1993siamese} to enforce the representation of microstructures in the latent feature space to preserve the microstructure distances in the original representation space.
% (aiming to find microstructures as close as possible to given properties)
% The corresponding\textbf{found oder identified? corresponding passt hier nicht} microstructures are then delivered\textbf{reconstructed?} via the decoder network. 

\subsubsection{Multi-task Learning (MTL)}
\label{subsec:Multi-task-learning}
\begin{figure}[h]
	\centering
  	\includegraphics[width=0.95\linewidth]{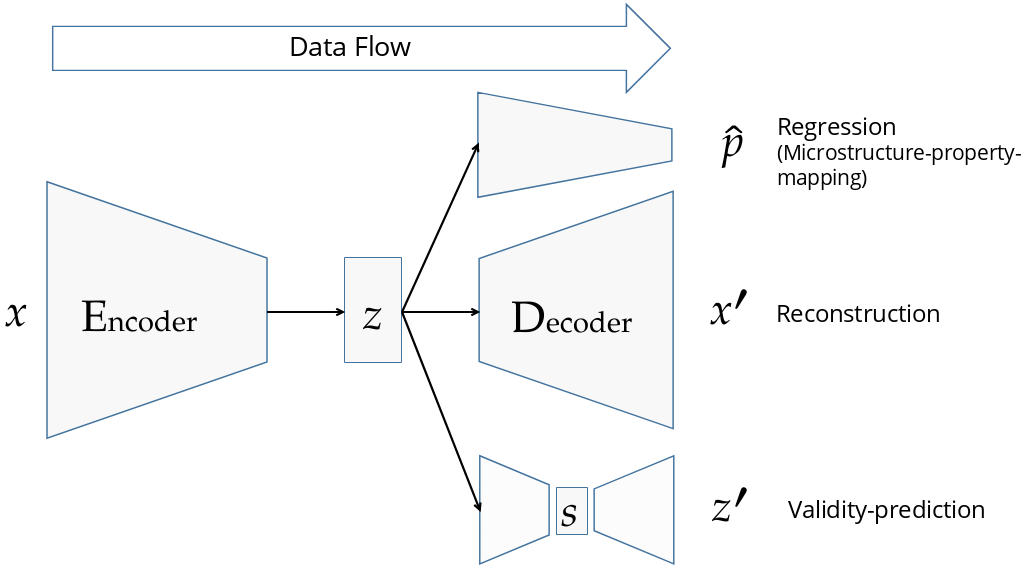}
\caption{MTL architecture. The MTL model is trained on pairs of microstructures and corresponding properties $(\boldsymbol{x},\boldsymbol{p})$. The input microstructures are transformed into latent features $\boldsymbol{z}$. The individual outputs of the connected tasks are the estimated properties $\boldsymbol{\hat{p}}$, the reconstructed microstructure $\boldsymbol{x}^\prime$ and the reconstructed latent features $\boldsymbol{z}^\prime$.}
	\label{img:Multi-task-learning-approach}
\end{figure}

The MTL processing scheme (shown in Fig. \ref{img:Multi-task-learning-approach}) starts with an encoder network which extracts significant features by mapping the microstructure space $\boldsymbol{x} \in \mathbb{R}^K$ into a lower dimensional latent feature space $\boldsymbol{z} \in \mathbb{R}^M$ via the learned function 
\begin{align}
\boldsymbol{z} = f_{\mathrm{enc}} (\boldsymbol{x}, \boldsymbol{\theta}_{\mathrm{enc}}),
\end{align}
in which the encoder network is parameterized by its weight values $\boldsymbol{\theta}_{\mathrm{enc}}$. All three previously described tasks are attached to the encoder in the form of feedforward neural networks. Besides, the encoder can be easily adapted to higher dimensional microstructure representing data types like images (EBSD or micrograph images) or three dimensional microstructure data by using for example convolutional neural networks (see \cite{Krizhevsky2012-CNN}), which are used for example in \cite{cecen2018material} in the materials sciences domain.
%The training input of the encoder network consist of $K$ sampled microstructures $\{\boldsymbol{x}_k\}_{k=1}^K$ and their corresponding properties $\{\boldsymbol{p}_k\}_{k=1}^K$. 

To train the MTL model, a loss function that combines all the three tasks is needed. This is achieved by a function that cumulates the loss terms of the three tasks $\mathscr{L}_{\mathrm{regr}}$ (regression loss), $\mathscr{L}_{\mathrm{recon}}$ (reconstruction loss) and $\mathscr{L}_{\mathrm{valid}}$ (validity loss), and weights them using $\mathscr{W}_{\mathrm{regr}}$, $\mathscr{W}_{\mathrm{recon}}$ and $\mathscr{W}_{\mathrm{valid}}$ to allow for prioritization. The total loss function is defined as
\begin{equation}\label{eq:mtl-loss} 
\begin{aligned}
	\mathscr{L}_{\mathrm{MTL}} = \mathscr{W}_{\mathrm{regr}} \mathscr{L}_{\mathrm{regr}} + \mathscr{W}_{\mathrm{recon}} \mathscr{L}_{\mathrm{recon}}  \\
						+ \mathscr{W}_{\mathrm{valid}} \mathscr{L}_{\mathrm{valid}} + \lambda R(\boldsymbol{\theta}),
	\end{aligned}
\end{equation} 
where $R(\boldsymbol{\theta})$ is a regularization term that is used to prevent overfitting with the hyperparameter $\lambda$ defining the strength of the regularization (also known as weight decay, see \cite{Krogh1991Weight-Decay} and \cite{Hinton1987Weight-Decay}). 
Each of the feedforward neural networks is parameterized by the respective weight values $\boldsymbol{\theta}_{\mathrm{enc}}$, $\boldsymbol{\theta}_{\mathrm{regr}}$, $\boldsymbol{\theta}_{\mathrm{recon}}$ and $\boldsymbol{\theta}_{\mathrm{valid}}$, which are adjusted simultaneously during training and altogether form the weight vector $\boldsymbol{\theta}$. In the following we will introduce the three individual loss terms.

\begin{enumerate}
\item The forward mapping of the latent feature vector $\boldsymbol{z}$ to the properties vector $\boldsymbol{\hat{p}} \in\mathbb{R}^N$ is represented by the learned function 
\begin{align}
\boldsymbol{\hat{p}} = f_{\mathrm{regr}} (\boldsymbol{z}, \boldsymbol{\theta}_{\mathrm{regr}}) = f_{\mathrm{regr}}(f_{\mathrm{enc}} (\boldsymbol{x}, \boldsymbol{\theta}_{\mathrm{enc}}), \boldsymbol{\theta}_{\mathrm{regr}}).
\end{align}
The regression loss is given by the mean squared error between the predicted properties $\boldsymbol{\hat{p}}$ and the true properties $\boldsymbol{p}$:
\begin{equation}\label{eq:regr-loss} 
	\mathscr{L}_{\mathrm{regr}} (\boldsymbol{p}, \boldsymbol{\hat{p}}) = \frac{1}{N} \sum_{i=1}^N ({p_i} - {\hat{p}_i} )^2,
\end{equation} 
where $N$ denotes the number of properties.

\item The decoder network, which is responsible for the reconstruction, transforms the latent feature vectors $\boldsymbol{z}$ back to the original microstructure space:
\begin{align}
\boldsymbol{x}^\prime = f_{\mathrm{recon}}(\boldsymbol{z}, \boldsymbol{\theta}_{\mathrm{recon}}) = f_{\mathrm{recon}}(f_{\mathrm{enc}}(\boldsymbol{x}, \boldsymbol{\theta}_{\mathrm{enc}}), \boldsymbol{\theta}_{\mathrm{recon}}).
\end{align}
The reconstruction loss is defined on the basis of a distance measure between two microstructural feature vectors $\text{dist}(\boldsymbol{x}, \boldsymbol{x^\prime} )$:
\begin{equation}\label{eq:recon-loss} 
	\mathscr{L}_{\mathrm{recon}} (\boldsymbol{x}, \boldsymbol{x^\prime}) = \text{dist}(\boldsymbol{x}, \boldsymbol{x^\prime} ).
\end{equation} 
The distance measure between depends on the microstructure representation and has to be chosen appropriately.

\item On the basis of the latent feature space, an extra autoencoder network is set up transforming $\boldsymbol{z} \in \mathbb{R}^M$ into an even lower-dimensional feature sub-space $\boldsymbol{s} \in \mathbb{R}^S$ with $S<M$ and transforming back to $\boldsymbol{z}^\prime \in \mathbb{R}^M$ via 
\begin{align}
\boldsymbol{z^\prime} = f_{\mathrm{valid}}(\boldsymbol{z}, \boldsymbol{\theta}_{\mathrm{valid}}) = f_{\mathrm{valid}}(f_{\mathrm{enc}}(\boldsymbol{x}, \boldsymbol{\theta}_{\mathrm{enc}}), \boldsymbol{\theta}_{\mathrm{valid}}).
\end{align}
The validity loss is defined by the mean squared error between $\boldsymbol{z}$ and $\boldsymbol{z}^\prime$:
\begin{equation}\label{eq:reach-loss} 
	\mathscr{L}_{\mathrm{valid}} (\boldsymbol{z}, \boldsymbol{z}^\prime) = \frac{1}{M} \sum_{i=1}^M ({z_i} - {z_i}^\prime )^2 .
\end{equation} 

\end{enumerate}

% Dedicated processes under consideration can only generate a restricted set of microstructures. The producibility of property-fitting microstructures found by the optimizer is an important information. Producible microstructures will only populate a sub-space or a sub-volume of the microstructure representation space, which can be found again by an auto-encoder, and where the distance between input and output indicates, how far the latent space representation is away from producible microstructures\textbf{diese information steht teilweise schon oben und sollte meiner meinung nach auch nur dort stehen. hier geht es rein um die definition des loss terms. es sollte bereits klar sein wieso wir dieses loss brauchen}. 

%\textbf{info: detailliertes feedback geht bis hier}

\subsubsection{Distance preserving feature extraction using siamese neural networks}
\label{subsec:Siamese-MTL}

The above described MTL approach is used in combination with an optimizer that searches for candidate microstructures with desired properties in the latent feature space. However, our approach aims to identify a diverse set of microstructures with high diversity. For the diversity quantification a distance measure in the latent feature space is required. The MTL approach as defined above, is not able to preserve the distances of the original space in the latent feature space. In order to construct a distance preserving latent feature space, the MTL is embedded in a siamese neural network \cite{bromley1993siamese,Chicco2021}, which we will describe next.

\begin{figure}[h]
	\centering
  	\includegraphics[width=0.95\linewidth]{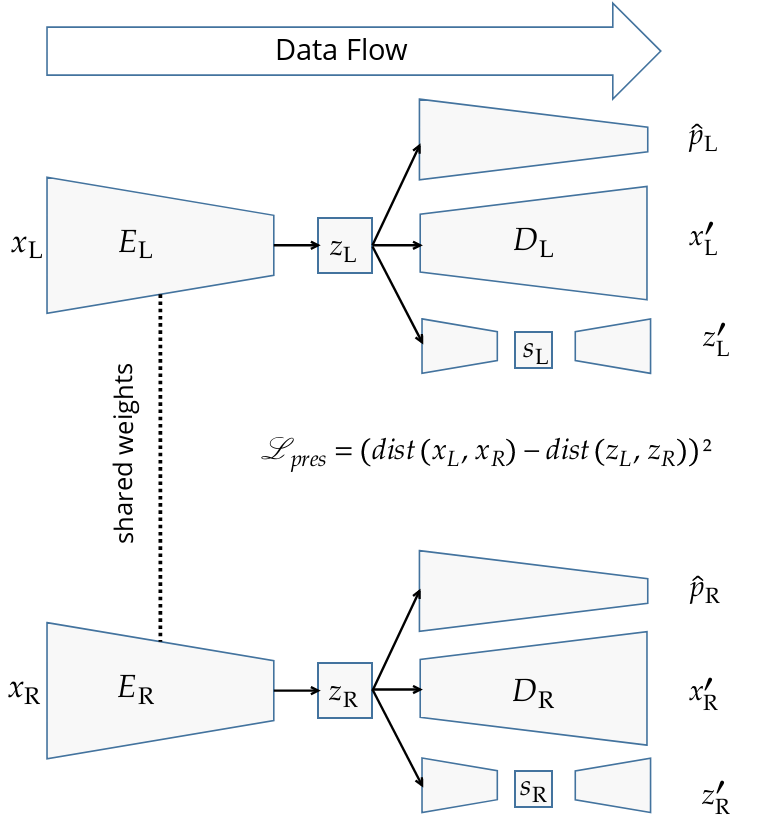}
	\caption{Architecture of the SMTL approach. The dotted line between the encoders $E_\mathrm{L}$ and $E_\mathrm{R}$ indicates shared weights.} 
	\label{img:Siamese-multi-task-learning}
\end{figure}

Siamese neural networks consist of two identical networks, which share weights in the encoder part, see Fig. \ref{img:Siamese-multi-task-learning}. Both networks embed different microstructures $\vec{x}_L$ and $\vec{x}_R$ as $\vec{z}_L$ and $\vec{z}_R$ in the latent feature space which is finally processed by two identical MTL networks. The distance preservation is enforced by defining a distance preservation loss $\mathscr{L}_{\mathrm{pres}}$ that minimizes the difference between the distance of two different input microstructures in the original space $\text{dist}(\boldsymbol{x}_L, \boldsymbol{x}_R)$ and the corresponding distance in the latent feature space $\text{dist}(\boldsymbol{z}_L, \boldsymbol{z}_R)$, with $\boldsymbol{x}_L \neq \boldsymbol{x}_R$ \cite{A-Siamese-Autoencoder-Preserving-Distances-2017}:
\begin{equation}\label{eq:pres-loss} 
	\mathscr{L}_{\mathrm{pres}}  = (\text{dist}(\boldsymbol{x}_L, \boldsymbol{x}_R) - \text{dist}(\boldsymbol{z}_L, \boldsymbol{z}_R))^2,
\end{equation} 
while $\text{dist}(\boldsymbol{x}_L, \boldsymbol{x}_R)$ and $ \text{dist}(\boldsymbol{z}_L, \boldsymbol{z}_R)$ are not necessarily the same distance measures. Applying such loss terms leads to multi dimensional scaling, see \cite{Kruskal1964MultidimensionalSB} and \cite{Multidimensional-Scaling-Cox2008}. Using the distance preservation loss $\mathscr{L}_{\mathrm{pres}}$, the MTL loss function, defined in Eq. \ref{eq:mtl-loss}, is extended by the weighted preservation loss $\mathscr{W}_{\mathrm{pres}} \mathscr{L}_{\mathrm{pres}}$ to
\begin{equation}\label{eq:siamese-mtl-loss} 
\begin{aligned}
	\mathscr{L}_{\mathrm{SMTL}} &= \mathscr{W}_{\mathrm{regr}} \mathscr{L}_{\mathrm{regr}}  + \mathscr{W}_{\mathrm{recon}} \mathscr{L}_{\mathrm{recon}} \\
	&+ \mathscr{W}_{\mathrm{valid}} \mathscr{L}_{\mathrm{valid}}  + 	\mathscr{W}_{\mathrm{pres}} \mathscr{L}_{\mathrm{pres}}  \\
	&+ \lambda R(\boldsymbol{\theta}).
\end{aligned}
\end{equation} 

The SMTL approach delivers a function which can map a microstructure representation in the latent feature space on properties. Now an optimizer can operate on a low dimensional feature space to find microstructures with desired properties. The SMTL framework also allows to reconstruct the original represenation of microstructures, to asses the distances between them and to validate them in the latent feature space.

%\textbf{info: schnelles feedback geht bis hier}

\subsubsection{Microstructure optimizer}
\label{subsec:Optimizer}

The microstructure optimization with respect to desired properties uses the distance preserving SMTL framework with the tasks microstructure-property-mapping, validity-prediction and reconstruction. The optimization minimizes a loss function, which consists of the cost terms $\mathscr{C}_\mathrm{prop}$, $\mathscr{C}_\mathrm{valid}$ and $\mathscr{C}_\mathrm{divers}$ and the corresponding weights $\mathscr{V}_\mathrm{prop}$, $\mathscr{V}_\mathrm{valid}$ and $\mathscr{V}_\mathrm{divers}$:
\begin{equation}\label{eq:score-fitness} 
	\mathscr{F} =  \mathscr{V}_\mathrm{prop} \mathscr{C}_\mathrm{prop} + \mathscr{V}_\mathrm{valid} \mathscr{C}_\mathrm{valid} + \mathscr{V}_\mathrm{divers} (1+\mathscr{C}_\mathrm{divers}).
\end{equation} 
$\mathscr{C}_\mathrm{prop}$, $\mathscr{C}_\mathrm{valid}$ and $\mathscr{C}_\mathrm{divers}$ denote the property, validity and diversity cost terms, respectively. While the property cost term drives the candidate microstructures to lie inside a specified target properties region, the validity cost aims that the optimizer operates inside the region of valid microstructures and the diversity cost ensures that candidate microstructures differ from each other. To minimize the loss function we use genetic algorithms, which generate a population set of $P$ candidate microstructures $\boldsymbol{\tilde{z}}^*$ in the latent feature space in every iteration. The three cost terms are described in more detail in the following.

\begin{enumerate}
\item The property cost is defined by the mean squared error between the desired properties and the predicted properties from the SMTL regression model:
\begin{equation}\label{eq:score-prop} 
	\mathscr{C}_\mathrm{prop} = \frac{1}{N} \sum_{i=1}^N (\widetilde{\mathscr{C}}_{\mathrm{prop},i} )^2.
\end{equation} 
If one of the predicted properties lies inside the target region, the cost $\widetilde{\mathscr{C}}_{\mathrm{prop},i}$ equals $0$. Otherwise, $\widetilde{\mathscr{C}}_{\mathrm{prop},i}$ equals the minimum squared distance from the predicted properties to the target region borders.

\item The validity prediction is used to asses whether an identified candidate microstructure is likely to be represented by the sample data set. The validity cost is defined by
\begin{equation}\label{eq:score-reach}  
	\mathscr{C}_\mathrm{valid} = \mathrm{max}(\mathscr{A} - \xi_\mathrm{valid}, 0),
\end{equation}
in which $\xi_\mathrm{valid}$ is a threshold to define the maximum tolerated reconstruction error for valid textures and $\mathscr{A}$ denotes the anomaly score
\begin{equation}\label{eq:anomaly-score}  
	\mathscr{A} = \frac{1}{M} \sum_{i=1}^M ({z^*_i} - z^{*\prime}_i )^2.
\end{equation}

%	\mathscr{A} = \frac{1}{M} \sum_{i=1}^M ({\boldsymbol{z}^*_i} - \boldsymbol{z}^{*\prime}_i )^2.

\item The diversity cost is based on the sum of the distances between the candidate microstructure $\boldsymbol{z}^*$ in the latent feature space and every other microstructure in the population:
\begin{equation}\label{eq:score-divers} 
	%\mathscr{S}_{divers} = - \sum_{i=1}^P \text{dist}(\boldsymbol{c}, P_i).
	\mathscr{C}_\mathrm{divers} = - \sum_{i=1}^P \text{dist}(\boldsymbol{z}^*_i, \boldsymbol{z}^*),
\end{equation} 
in which for $\text{dist}(\boldsymbol{z}^*_i, \boldsymbol{z}^*)$ the same distance measure has to be used as for the latent feature vectors in Eq. \ref{eq:pres-loss}. %In Eq. \ref{eq:score-divers}, a low sum of pairwise distances means a high similarity between the individuals and therefore a low diversity, whereas a high sum means a low similarity and therefore high diversity.
\end{enumerate}

\subsection{Materials science fundamentals}
\label{sec:methods_material_modelling}

\subsubsection{Representation of crystallographic texture}

Crystallographic texture is typically described by the orientation distribution function, which is defined by
\begin{equation}
f(g)\mathrm{d}g = \frac{V(g)}{V},
\end{equation}
for an orientation $g$ (a point in $SO(3)$) and the volume $V(g)$ in $SO(3)$. 
%The orientation distribution function is continuous, non-negative and the total integral of it equals one. 
The orientation distribution function $f(g)$ often underlies specific symmetry conditions, for which various regions in $SO(3)$ are equivalent. Therefore, depending on the symmetries, orientations can be mapped into an elementary region of $SO(3)$, the so-called fundamental zone. The orientation distribution function on the basis of the orientations mapped into the fundamental zone is then indistinguishable from the original orientation distribution function. Rolling textures, for example, underlie a cubic crystal and an orthorhombic sample symmetry, for which 96 elementary regions exist \cite{hansen1978tables}.

A popular way to represent the orientation distribution function is by approximating it via generalized spherical harmonic functions \cite{bunge2013texture}. Yet, as there is no straightforward way to measure the distance between two orientation distribution functions in terms of generalized spherical harmonics, we make use of the orientation histogram-based texture descriptor, which is introduced in \cite{dornheim2020structureguided}. 
Therefore, the cubic fundamental zone is discretized into a set $O$ of $j$ nearly uniform distributed orientations $o_j$. For each individual orientation $g$ in a set of orientations $G$, a weight vector $w_\mathrm{g}$ is constructed via a soft-assignment approach
\begin{equation}
w_\mathrm{g} = 
\left\{\begin{array}{ll}
\frac{\Phi(g,o_j)}{\sum_{o_i \in N_l} \Phi(g,o_i)}, & \text{if} ~~ o_j \in N_l \\
0, & \text{else} 
\end{array}\right. ,
\end{equation}
where $N_l$ is the set of $l$ nearest neighbor orientations of  $g$ in terms of the orientation distance $\Phi$. The orientation distance between two orientations $g$ and $o$ is defined by
\begin{equation}
\Phi = \min \Phi(\overline{g},\overline{o}).
\end{equation} 
where $\overline{g}$ and $\overline{o}$ is from the set of all equivalent orientations of $g$ and $o$ in terms of cubic crystal symmetry. The orientation distance measure in $SO(3)$ is defined as
\begin{equation}
\Phi(q_\mathrm{g},q_\mathrm{o}) = \min(||q_\mathrm{g}-q_\mathrm{o}||,||q_\mathrm{g}+q_\mathrm{o}||),
\end{equation}
where $q_\mathrm{g}$ and $q_\mathrm{o}$ are the quaternion representations of the orientations $g$ and $o$ \cite{Huynh2009}. 

On this basis, the weight vector for the orientation histogram $\boldsymbol{b}$ can be calculated by a volume average of the weight vectors of the individual orientations
\begin{equation}
\boldsymbol{b} = \frac{1}{V}\sum_j V(o_j)w_{o_j}.
\end{equation}
The distance between two orientation distribution functions can then be measured via any kind of histogram-based distance measure, such as the Chi-Squared distance \cite{Pele2010chi-square-distance}
\begin{equation}
\chi^2(\boldsymbol{b}_1,\boldsymbol{b}_2) = \sum \frac{(\boldsymbol{b}_1-\boldsymbol{b}_2)^2}{\boldsymbol{b}_1+\boldsymbol{b}_2}.
\label{eq:chi2_whd}
\end{equation}

The set of nearly uniform distributed orientations $O$, needed for the histogram-based texture descriptor, can be generated using the algorithm described in \cite{Quey2018}, which is implemented in the software \textit{neper} \cite{quey2011large}. For the purpose of this study, we sample $512$ nearly uniform distributed orientations over the cubic fundamental zone and chose a soft assignment of $l=3$.

\subsubsection{Crystallographic texture of steel sheets}

After rolling body centered cubic (bcc) materials, typically so-called fiber textures are formed. Following \cite{ray1994cold}, these textures are composed of the five fibers $\alpha$, $\gamma$, $\eta$, $\epsilon$, and $\beta$, which are defined in detail in Tab. \ref{tab:fibers_definition}. Among these fibers, the $\alpha$ and $\gamma$ fiber are most prominent \cite{kocks1998texture}, whereas the presence of the $\beta$ fiber is only reported from theoretical predictions \cite{von1986investigation}.
\begin{table}
	\caption{Definition of the fibers of bcc rolling textures following \cite{kocks1998texture}}
	\label{tab:fibers_definition}
	\begin{center}
		\begin{tabular}{| l | l |}
			\hline
			Fiber & Location \\
			\hline
			\hline
			$\alpha$ & from \{001\}$<$110$>$ to \{111\}$<$1$\overline{1}$0$>$, parallel to RD \\
			$\gamma$ & from \{111\}$<$1$\overline{1}$0$>$ to \{111\}$<$112$>$, parallel to ND \\
			$\eta$ & from \{001\}$<$100$>$ to \{011\}$<$100$>$, parallel to RD \\
			$\epsilon$ & from \{001\}$<$110$>$ to \{111\}$<$112$>$, parallel to TD \\
			$\beta$ & from \{112\}$<$1$\overline{1}$0$>$ to \{$\overline{11}$ $\overline{11}$ 8\}$<$4 4 $\overline{11}$ $>$  \\
			\hline
		\end{tabular}
	\end{center}
\end{table}
In order to generate a data base of (artificial) rolling textures, in this work, a $25$-parameter model is used, as it is proposed in \cite{delannay1999new} to describe steel sheet textures. The model is based on textures that are composed of the fibers $\alpha$, $\gamma$, and $\eta$. 

As the $\eta$-fiber is not always present in steel sheet textures, we limit ourselves to textures that consist of an $\alpha$ and $\gamma$ fiber. Therefore, $6$ of the $25$ parameters can be neglected. The texture model describes the orientation distribution function as a set of weighted Gaussian distributions placed along the fibers. The model parameters $D_i$ are listed in Tab. \ref{tab:delannay_parameter_model} and define the standard deviations and the mean values of the distributions based on the fiber thickness and the shifts from their ideal positions. Furthermore, the model parameters define the weights of the distributions among each other based on the fiber intensity.

\begin{table}
	\caption{Definition of the parameters $D_i$ of the texture model, cf. \cite{delannay1999new}. The ideal position is given in Bunge euler angles [$^\circ$].}
	\label{tab:delannay_parameter_model}
	\begin{center}
		\begin{tabular}{| l | c | c | c | c | c | c | c |}
			\hline
			 & Ideal pos. & Intens. & Std $\varphi_2$ & Std $\varphi_1$ & Shift $\varphi_1$ & Shift $\phi$ & Shift $\varphi_2$ \\
			\hline
			\hline
			$a_1$     & $0,0,45$   & $D_1$ & $D_7$    & $D_7$    & 0        & 0        & 0 \\
			$a_2$     & $0,30,45$  & $D_2$ & $D_8$    & $D_{13}$ & 0        & $D_{15}$ & 0 \\
			$a_3$ & $0,55,45$  & $D_3$ & $D_9$    & $D_9$    & 0        & $D_{16}$ & 0 \\
			$a_4$     & $0,70,45$  & $D_4$ & $D_{10}$ & $D_{10}$ & 0        & $D_{17}$ & 0 \\
			$a_5$     & $0,90,45$  & $D_4$ & $D_{10}$ & $D_{10}$ & 0        & 0        & 0 \\
			$g_2$     & $15,55,45$ & $D_5$ & $D_{11}$ & $D_{11}$ & $D_{14}$ & 0        & 0 \\
			$g_3$     & $30,55,45$ & $D_6$ & $D_{12}$ & $D_{12}$ & 0        & $D_{18}$ & $D_{19}$ \\
			\hline
		\end{tabular}
	\end{center}
\end{table}

To construct the set of Gaussian distributions, the seven base distributions from Tab. \ref{tab:delannay_parameter_model} are placed at their ideal positions with respect to the shifts. Between these seven distributions, further distributions are placed with a distance of about $3^\circ$ to each other, leading to overall $41$ Gaussians. Their weights $w_i$ and the values for the standard deviation $\sigma_i$ and mean value $\mu_i$ are interpolated linearly based on the values of the two neighboring base distributions. This yields a set of Gaussian distributions $\mathcal{N}_1(\mu_1,\sigma_1), ..., \mathcal{N}_{41}(\mu_{41},\sigma_{41})$. The orientation distribution function $f(g)$ is defined by the normalized sum of this set:
\begin{equation}
f(g) = \frac{1}{\sum_i w_i} \sum_{i=1}^{n} w_i \mathcal{N}_i(\mu_i,\sigma_i).
%f(g) = \sum_{i=1}^{n} w_i \mathcal{N}_i(\mu_i,\sigma_i), ~~~\sum_i w_i=1.
\label{eq:odf}
\end{equation}

Based on this definition, discrete orientations can be sampled. In the following, we denote the set of orientations as $G$. As $f(g)$ is defined in the cubic-orthorhombic fundamental zone, it is necessary to add the equivalent orientations regarding the orthorhombic sample symmetry to the set of discrete orientations. This is done by applying rotation operations $g_s$ on each orientation $g_i$ in $G$
\begin{equation}
g_i^\mathrm{equiv} = g_s ~ g_i.
\end{equation}
The rotation operations $g_s$ for orthorhombic sample symmetry can be found in \cite{hansen1978tables}.

\subsubsection{Material model}

The sheet metal properties which we focus on in this study are the Young's moduli and the R-values at $0$, $45$ and $90$ degree to rolling direction. In this study, the properties are calculated by applying uniaxial tension on a crystal plasticity-based material model. As time efficiency is essential for the generation of data, a material model of Taylor-type is implemented, as it is described in \cite{dornheim2020structureguided}.

The Taylor-type material model is based on the volume averaged stress of a set of $n$ crystals \cite{Kalidindi.1992}:
\begin{equation}
\boldsymbol{\overline{T}} = \frac{1}{V}\sum_{i=1}^n \boldsymbol{T}^{(i)}V^{(i)}.
\end{equation}
In the above equation, $\boldsymbol{T}$ denotes the Cauchy stress tensor, which can be derived by the stress tensor in the intermediate configuration, given by 
\begin{equation}
%\boldsymbol{T}^*=\frac{1}{2} ~ \mathbb{C}:(\boldsymbol{C}_\mathrm{e}-\boldsymbol{I})=\frac{1}{2} ~ \mathbb{C}:(\boldsymbol{F}_\mathrm{e}^T\cdot\boldsymbol{F}_\mathrm{e}-\boldsymbol{I}),
\boldsymbol{T}^*=\frac{1}{2} ~ \mathbb{C}:(\boldsymbol{F}_\mathrm{e}^T\cdot\boldsymbol{F}_\mathrm{e}-\boldsymbol{I}),
\end{equation}
with the second order identity tensor $\boldsymbol{I}$ and the fourth order elastic stiffness tensor $\mathbb{C}$. The elastic constants $C_{11}$, $C_{12}$ and $C_{44}$ are set to $218.37$, $131.13$ and $105.34$ GPa, respectively \cite{eghtesad2020high}. $\boldsymbol{F}_\mathrm{e}$ is the elastic part of the deformation gradient $\boldsymbol{F}$ and can be calculated by a multiplicative decomposition
\begin{equation}
\boldsymbol{F} = \boldsymbol{F}_\mathrm{e} \cdot \boldsymbol{F}_\mathrm{p}.
\end{equation}

The intermediate stress tensor can be converted into Cauchy stress using the relation
\begin{equation}
\boldsymbol{T^*} = \boldsymbol{F}_\mathrm{e}^{-1}\cdot(\operatorname{det}(\boldsymbol{F}_\mathrm{e})~\boldsymbol{T})\cdot\boldsymbol{F}_\mathrm{e}^{-\top}.
\end{equation}
To describe the evolution of the plastic deformation, the plastic part of the velocity gradient $\boldsymbol{L}_\mathrm{p}$ is considered by
\begin{equation}
\boldsymbol{L}_\mathrm{p} = \dot{\boldsymbol{F}}_\mathrm{p} \cdot\boldsymbol{F}_\mathrm{p}^{-1},
\end{equation}
and the flow rule \cite{Rice.1971}
\begin{equation}
\boldsymbol{L}_\mathrm{p}=\sum_\eta \dot{\gamma}^{(\eta)} \boldsymbol{m}^{(\eta)} \otimes \boldsymbol{n}^{(\eta)},
\end{equation}
where $\dot{\gamma}^{(\eta)}$ denotes the shear rates on the active slip systems $\eta$, defined by the slip plane normal $\boldsymbol{n}^{(\eta)}$ and the slip direction $\boldsymbol{m}^{(\eta)}$. For bcc materials, the slip system families in terms of Miller index are \{110\}$<$111$>$, \{112\}$<$111$>$, and \{123\}$<$111$>$, while the latter is neglected due to simplicity.

The shear rates are defined by a phenomenological power-law \cite{Asaro.1985}:
\begin{equation}
\dot{\gamma}^{(\eta)}=\dot{\gamma}_0 \left| \frac{\tau^{(\eta)}}{r^{(\eta)}} \right|^{1/m}\mathrm{sign}(\tau^{(\eta)}),
\end{equation}
where $r^{(\eta)}$ is the slip system resistance, $\dot{\gamma}_0$ the reference shear rate and $m$ the shear rate sensitivity. Here, $\dot{\gamma}_0$ and $m$ are set to $0.001$ sec$^{-1}$ and $0.0125$, respectively, \cite{Pagenkopf.2016}.
Following Schmid's law, the resolved shear stress on slip system $\tau^{(\eta)}$ is given by
\begin{equation}
%\tau^{(\eta)}=((\boldsymbol{C}_\mathrm{e}\cdot\boldsymbol{T}^*):(\boldsymbol{m}^{(\eta)}\otimes\boldsymbol{n}^{(\eta)}).
\tau^{(\eta)}=((\boldsymbol{F}_\mathrm{e}^T\cdot\boldsymbol{F}_\mathrm{e})\cdot\boldsymbol{T}^*):(\boldsymbol{m}^{(\eta)}\otimes\boldsymbol{n}^{(\eta)}),
\end{equation}
and the evolution of the slip system resistance is defined by
\begin{equation}
\dot{r}^{(\eta)} = \frac{\mathrm{d}\hat{\tau}^{(\eta)}}{\mathrm{d}\Gamma}\sum_\xi q_{\eta\xi}|\dot{\gamma}^{(\xi)}|.
\end{equation}

The matrix $q_{\eta\xi}$ describes the ratio between self and latent hardening. It consists of diagonal elements equal to $1.0$ and off-diagonal elements $q_1$ and $q_2$, cf. \cite{baiker2014determination}. Both, $q_1$ and $q_2$, are set to $1.4$  \cite{Asaro.1985}. Further, we need to model the hardening behavior, which is realized by an extended Voce-type model \cite{Tome.1984}:
\begin{equation}
\hat{\tau}^{(\eta)}=\tau_0+(\tau_1+\vartheta_1\Gamma)(1-e^{-\Gamma\vartheta_0/\tau_1}).
\end{equation}
The material dependent parameters are calibrated to DC04 steel\footnote{experiments performed at IUL Dortmund during DFG project Graduate School 1483 \cite{Pagenkopf.2019}} and are $\tau_0=94.9$ MPa, $\tau_1=50$ MPa, $\vartheta_0=258$ MPa and $\vartheta_1=32.8$ MPa \cite{Pagenkopf.2019}. The accumulated plastic shear is defined by
\begin{equation}
\Gamma = \int_0^t \sum_\eta \left| \dot{\gamma}^{(\eta)} \right| \mathrm{d}t.
\end{equation}

\section{Results}
\label{sec:results}

\subsection{Texture-property data set}
\label{subsec:Data-generation}
For training, $50000$ sets of $2000$ discrete orientations are sampled via Latin Hypercube Design \cite{McKay.1979}, based on Eq. \ref{eq:odf}. In order to have an independent test set, further $10000$ sets are generated randomly. The ranges inside which the parameters of the texture model vary are defined such that typical bcc rolling textures found in literature can be represented, cf. \cite{lucke1970rolling,inagaki1972development,klinkenberg1992effects,kocks1998texture,kestens2016texture,Pagenkopf.2016,das2017calculation}. The parameter ranges are listed in Tab. \ref{tab:delannay_parameter_ranges}. In addition, to evaluate the anomaly detection, a set of artificial textures is needed, which slightly differ from the generated rolling textures. For this purpose, $10000$ anomalies are generated by shifting the $\alpha$-fiber (i.e. the ideal position of $a_1$, $a_2$, $a_4$ and $a_5$) about $20$ degrees in $\varphi_1$-direction.

\begin{table}
	\caption{Parameter ranges for $D_i$}
	\label{tab:delannay_parameter_ranges}
	\begin{center}
		\begin{tabular}{| l | c | c | c | c | c | c | c |}
			\hline
			Intensity & $D_1$ & $D_2$ & $D_3$ & $D_4$ & $D_5$ & $D_6$ & \\
			\hline
			\hline
			min & $1/6$ & $1/6$ & $1/6$ & $0$ & $1/6$ & $1/6$ & \\
			max & $1/3$ & $2/3$ & $2/3$ & $1/3$ & $1$ & $1$ & \\
			\hline
			\hline
			Std & $D_7$ & $D_8$ & $D_9$ & $D_{10}$ & $D_{11}$ & $D_{12}$ & $D_{13}$ \\
			\hline
			\hline
			min & $5/3$ & $5/3$ & $5/3$ & $5/3$ & $5/3$ & $5/3$ & $5/3$ \\
			max & $45/3$ & $35/3$ & $35/3$ & $30/3$ & $30/3$ & $30/3$ & $35/3$ \\
			\hline
			\hline
			Shift & $D_{14}$ & $D_{15}$ & $D_{16}$ & $D_{17}$ & $D_{18}$ & $D_{19}$ & \\
			\hline
			\hline
			min & $-5$ & $-10$ & $-10$ & $-5$ & $-10$ & $-10$ & \\
			max & $10$ & $10$ & $5$ & $10$ & $10$ & $10$ &  \\
			\hline
		\end{tabular}
	\end{center}
\end{table}

\begin{figure}
	\centering
  	\includegraphics[width=0.90\linewidth]{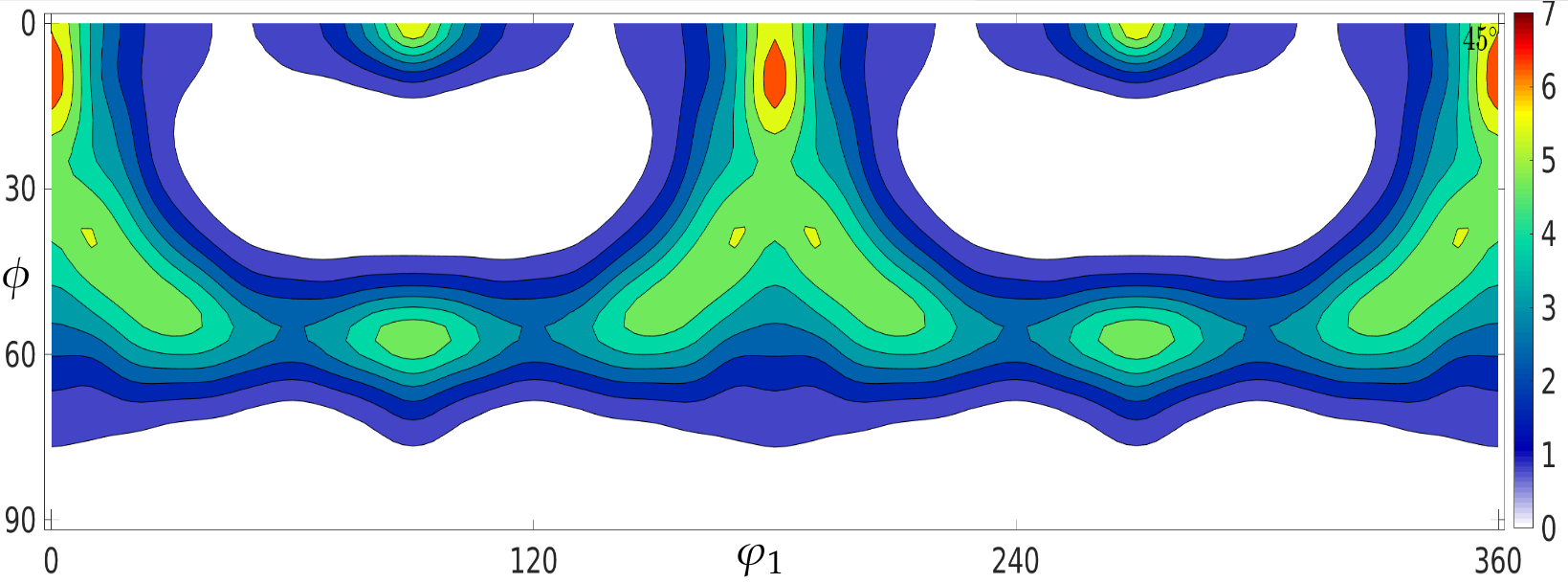}
	%\caption{Intersection of the euler space at $\varphi_2=45^\circ$ to visualize the $\gamma$-fiber of the colled rolled DC04 steel texture}.
	\caption{$\varphi_2=45^\circ$ section of the orientation distribution function to visualize the $\gamma$-fiber of the colled rolled DC04 steel texture}.
	\label{fig:dc04_texture_pole}
\end{figure}

Moreover, we want to validate the texture-property-mapping and the validity-prediction on experimental data. For this purpose, an experimentally measured texture of cold rolled DC04 steel from \cite{Schreijag.2012} is used. Based on this measurement, an orientation distribution function is approximated via the MATLAB toolbox mtex \cite{bachmann2010}, rotated into its symmetry axis assuming orthorhombic sample symmetry and mirrored.
To visualize the $\alpha$- and $\gamma$-fiber of the orientation distribution, an intersection plot of the euler space at $\varphi_2=45^\circ$ is depicted in Fig. \ref{fig:dc04_texture_pole}. 

\subsection{Validation of SMTL}
\label{sec:results_validation}
\begin{figure}
	\centering
  	\includegraphics[width=0.95\linewidth]{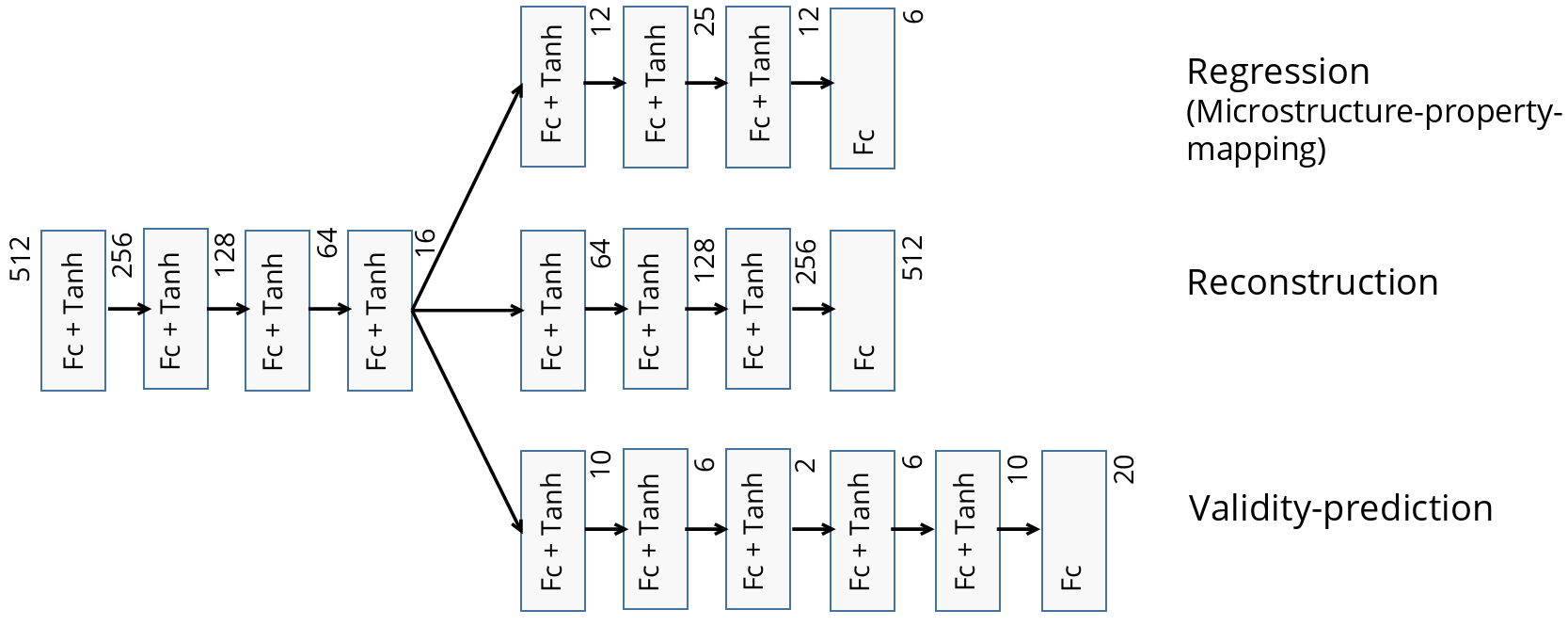}
	\caption{One twin part of the SMTL model with the annotation of the dimension size of the layers. $Fc$ denotes fully-connected layers and \textit{tanh} denotes hyperbolic tangent activation function}
	\label{fig:Model-Layers}
\end{figure}

In this study, the individual tasks of the SMTL model are realized via feedforward neural networks with \textit{tanh} activation functions to obtain features between $-1$ and $+1$ in the latent feature space. The SMTL model is implemented based on the Python TensorFlow API \cite{tensorflow2015-whitepaper}. The base network of the siamese architecture is illustrated in Fig. \ref{fig:Model-Layers}. The Glorot Normal method \cite{Glorot2010WeightInit} is used for weight initialization. In order to adjust the hyperparameters, a random search method \cite{Bergstra2012GridSearch} is applied using $5$-fold cross-validation. The best model configuration that was found is shown in Tab. \ref{tab:Hyperparameters}. We use the Chi-Squared distance introduced in Eq. \ref{eq:chi2_whd} as the distance measure in the input space. In the latent  latent feature space we use the sum of squared errors (SSE) between two vectors $\boldsymbol{z}_{\mathrm{1}}$ and $\boldsymbol{z}_{\mathrm{2}}$ as distance measure 
\begin{equation}\label{eq:distance-SSE}  
	\mathrm{SSE} (\boldsymbol{z}_{\mathrm{1}}, \boldsymbol{z}_{\mathrm{2}}) =  \sum_{i=1}^M  (\boldsymbol{z}_{\mathrm{1},i} - \boldsymbol{z}_{\mathrm{2},i})^2 .
\end{equation}
The SMTL model is trained for $200$ epochs, while the best intermediate result of the test set is retained, which can be interpreted as a form of early stopping \cite{Prechelt2012EarlyStopping}. Before the model training is executed, the loss terms are scaled to values between 0 and 1 in order to make them comparable. The following weights for the scaled loss terms were found to be appropriate from hyper parameter optimization: $\mathscr{W}_\mathrm{regr} = 0.05$, $\mathscr{W}_\mathrm{recon} = 0.05$, $\mathscr{W}_\mathrm{valid} = 0.05$ and $\mathscr{W}_\mathrm{pres} = 0.85$. 
\begin{table}
	\begin{center}
		\caption{Used hyperparamters}
		\label{tab:Hyperparameters}
		\begin{tabular}{| r | r | }
			\hline		
			Hyperparameter & Value 	\\
			\hline
			\hline	
			Optimizer 		& Adam \cite{Kingma2015Adam}	\\
			Learning rate 	& 0.001									\\
			Weight decay 	& 1e-6									\\
			Batch size 		& 256									\\
			\hline			 
		\end{tabular}
	\end{center}
\end{table}

The results for the texture-property-mapping and the distance preservation are shown in Tab. \ref{tab:Results-Siamese-Experimental}, in which the regression errors MAE$_\mathrm{E}$ and MAE$_\mathrm{r}$ denote the mean absolute error between the true and predicted Young's moduli and R-values depending on the dimension of the latent feature space $\boldsymbol{z}$. The quality of the distance preservation is measured by the coefficient of determination $R^2$, between the distances of two input textures and their corresponding latent feature vectors $R^2( \chi^2(\boldsymbol{x}_L, \boldsymbol{x}_R), SSE(\boldsymbol{z}_L, \boldsymbol{z}_R) )$. It is shown that texture-property-mappings with an adequate prediction quality can be achieved by extensively reducing the dimensionality of the latent feature space. However, regarding the distance preservation quality, a lower bound of at least $10$ latent features can be identified, below which the distance preservation is unsatisfactory. 
Additionally, the texture-property-mapping is evaluated on the experimentally measured texture and the corresponding properties. The results are listed in Tab. \ref{tab:Results-Siamese-Experimental}. It can be seen that a satisfactory prediction quality (Regr. MAE$_\mathrm{E} \leq 1000$ MPa and Regr. MAE$_\mathrm{r} \leq 0.1$) can only be achieved for at least $16$ latent features. %\textbf{TODO: MAE schraeg?}

\begin{table}
	\caption{Results for varying numbers of latent features (LF) of the texture-property-mapping and the distance preservation applied to the artificially generated textures and experimentally measured texture. Regr.(ession error) MAE$_\mathrm{E}$ is given in $[$MPa$]$, Regr.(ession error) MAE$_\mathrm{r}$ in $[$-$]$ and Pres.(erve quality) $R^2$ in $[$\%$]$.}
  \begin{tabular}{|c|c|c|c|c|c|}
    \hline
    \multirow{2}{*}{LF} &
      \multicolumn{3}{c|}{Artificial textures} &
      \multicolumn{2}{c|}{Experimental texture} \\

    & Regr. MAE$_\mathrm{E}$  & Regr. MAE$_\mathrm{r}$ & Pres. $R^2$  & Regr. MAE$_\mathrm{E}$ & Regr. MAE$_\mathrm{r}$ \\
    \hline
    \hline
	20 & 153					& 0.03						& 98.2 			& 596						& 0.04		\\
	18 & 183					& 0.03						& 98.0 			& 773						& 0.11		\\
	16 & 162					& 0.03						& 97.8 			& 907						& 0.05		\\
	14 & 193					& 0.03						& 97.0 			& 1282						& 0.07		\\
	12 & 215					& 0.04						& 95.4 			& 1468						& 0.13		\\
	10 & 238					& 0.05						& 92.2 			& 1463						& 0.13		\\
	 8 & 335					& 0.06						& 85.7 			& 1575						& 0.17		\\
	 6 & 390					& 0.07						& 72.4			& 1554						& 0.15		\\
	 4 & 664					& 0.10						& 34.2 			& 2768						& 0.12		\\
    \hline
  \end{tabular}
  \label{tab:Results-Siamese-Experimental}
\end{table}

\begin{figure}
	\centering
  	\includegraphics[width=0.95\linewidth]{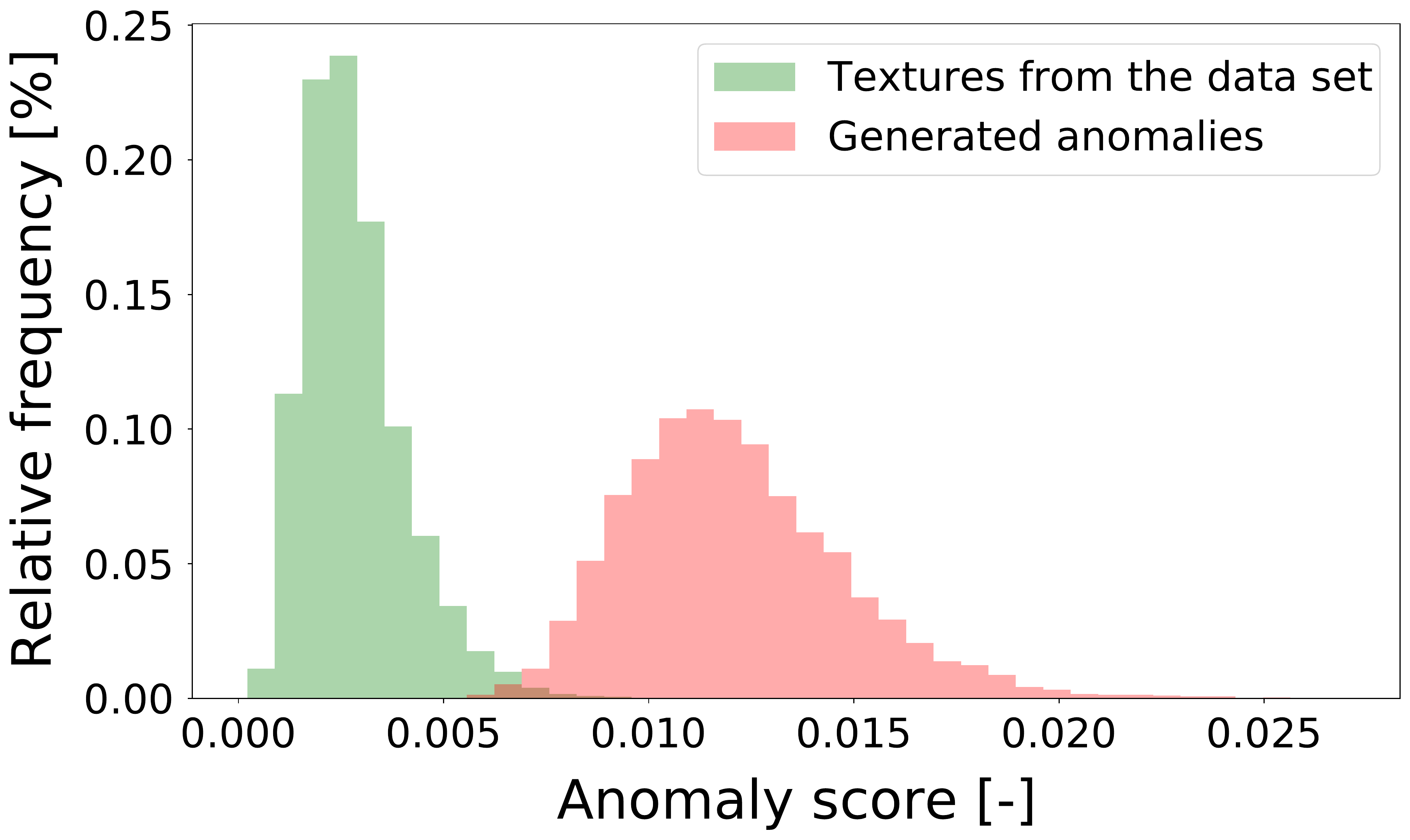}
	\caption{Histograms of the anomaly scores for the data from the test set and the set of artificially generated anomalies. The anomaly scores are based on the model that uses $16$ latent features.}
	\label{fig:Anomalie-detection_histogram}
\end{figure}

On the basis of this $16$-dimensional feature space, the validity-prediction is then evaluated. The anomaly scores for the textures in the test set and for the artificially generated anomalies are shown in Fig. \ref{fig:Anomalie-detection_histogram}. It can be seen, that the anomalies can be separated in a sufficient manner from the textures in the test set. 

%However, the anomaly score for the experimentally measured texture equals $0.0099$ and is therefore located in the transition zone between the textures from the data set and the anomalies. However, at this point we need to remark that it is the aim for the generated anomalies to be close to the generated rolling textures. Very different textures achieve comparably high anomaly scores.

\subsection{Rolling texture identification} 
\begin{figure}
  \centering
  \subfloat[][]{\includegraphics[width=0.50\textwidth]{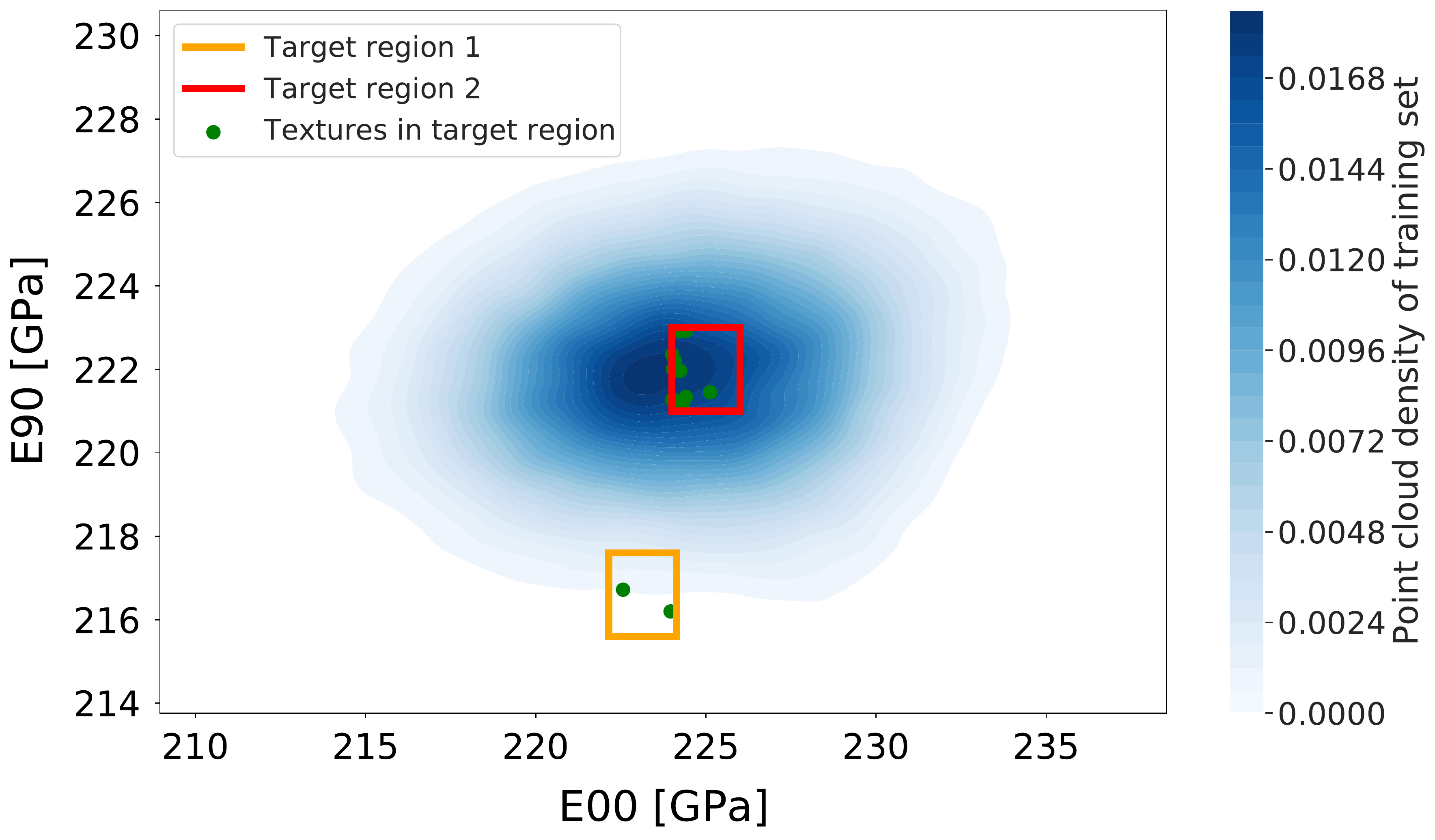}}%
  %\qquad
  \subfloat[][]{\includegraphics[width=0.50\textwidth]{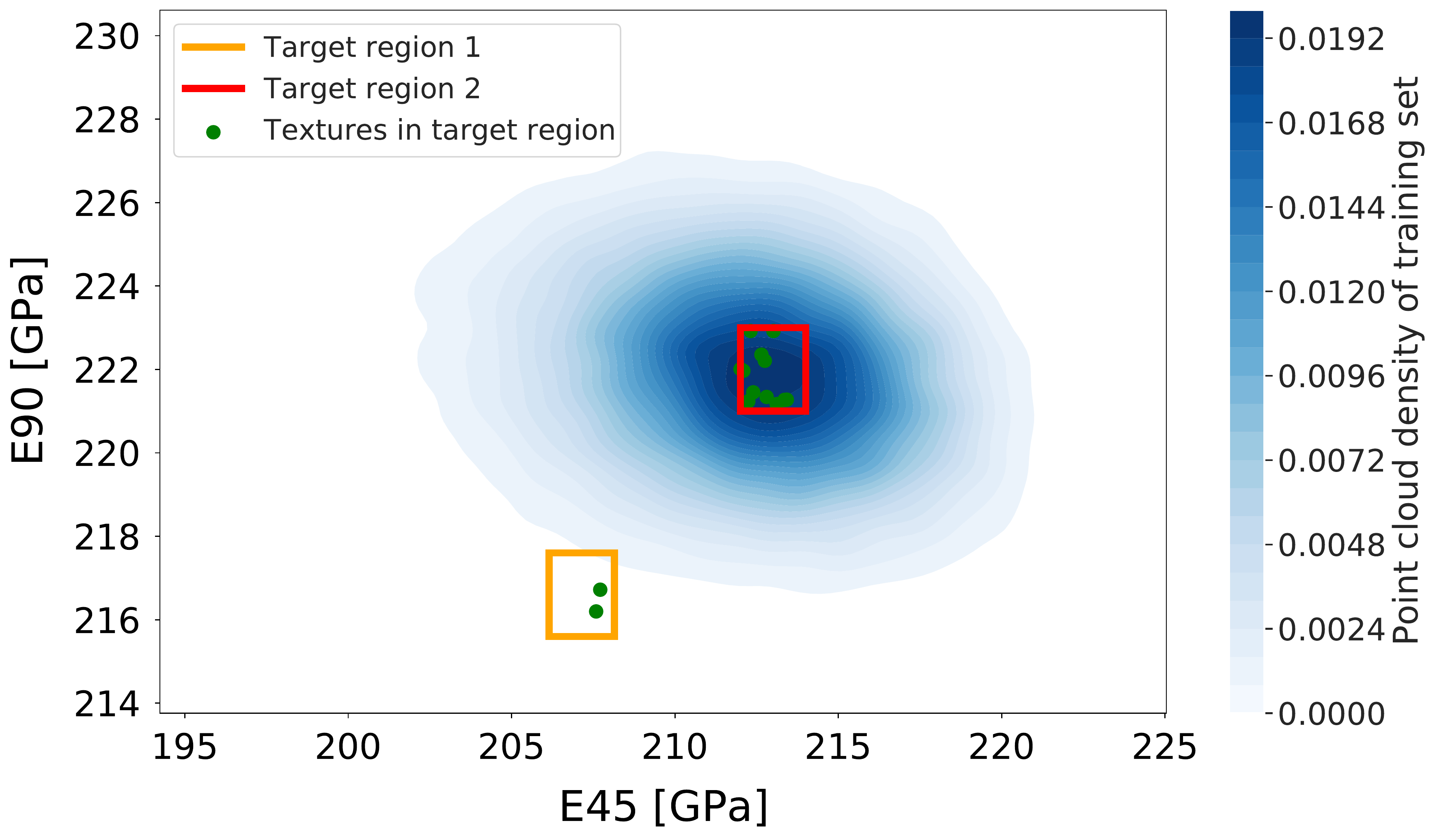}}%
  \qquad
  \subfloat[][]{\includegraphics[width=0.50\textwidth]{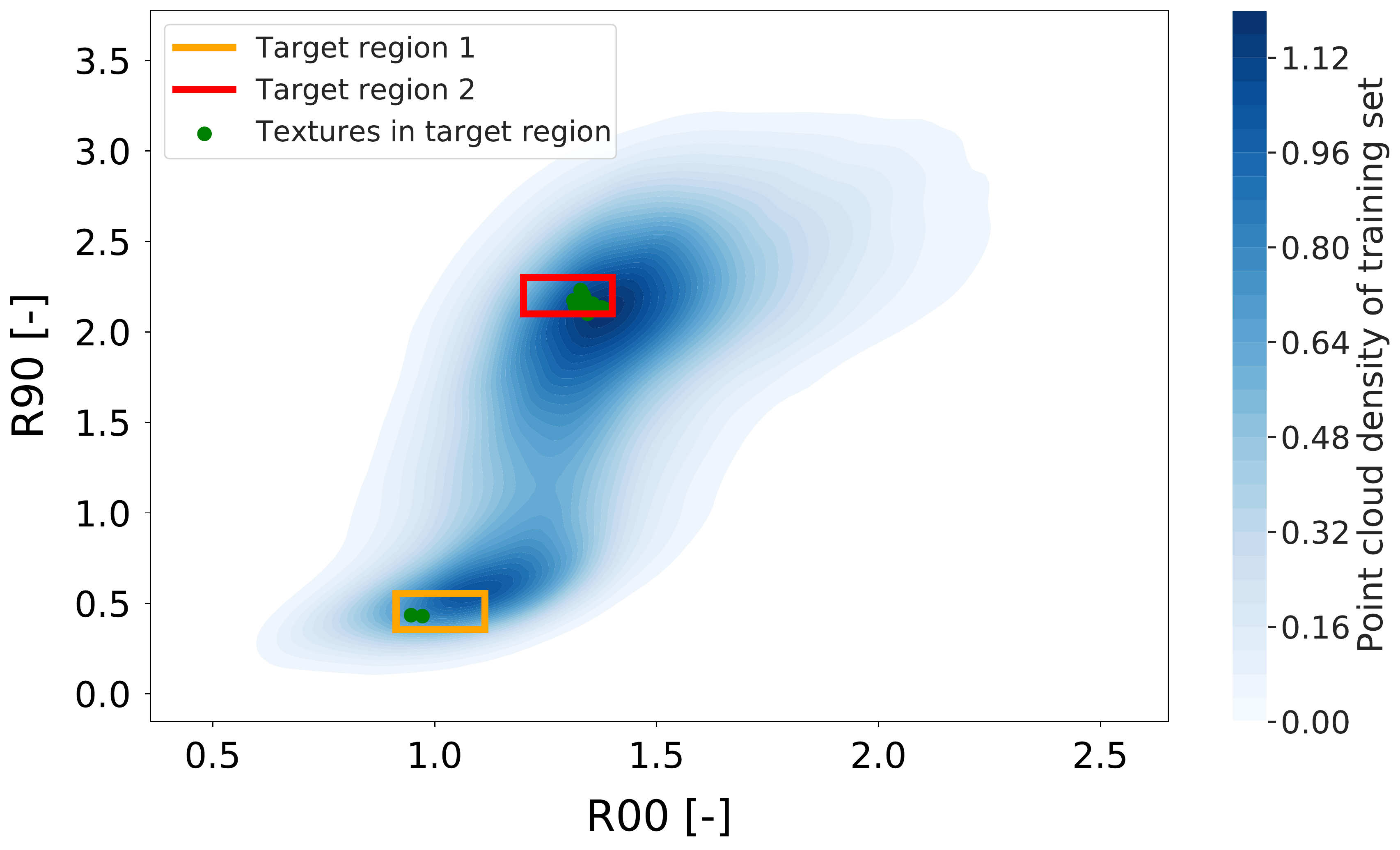}}%  
  %\qquad
  \subfloat[][]{\includegraphics[width=0.50\textwidth]{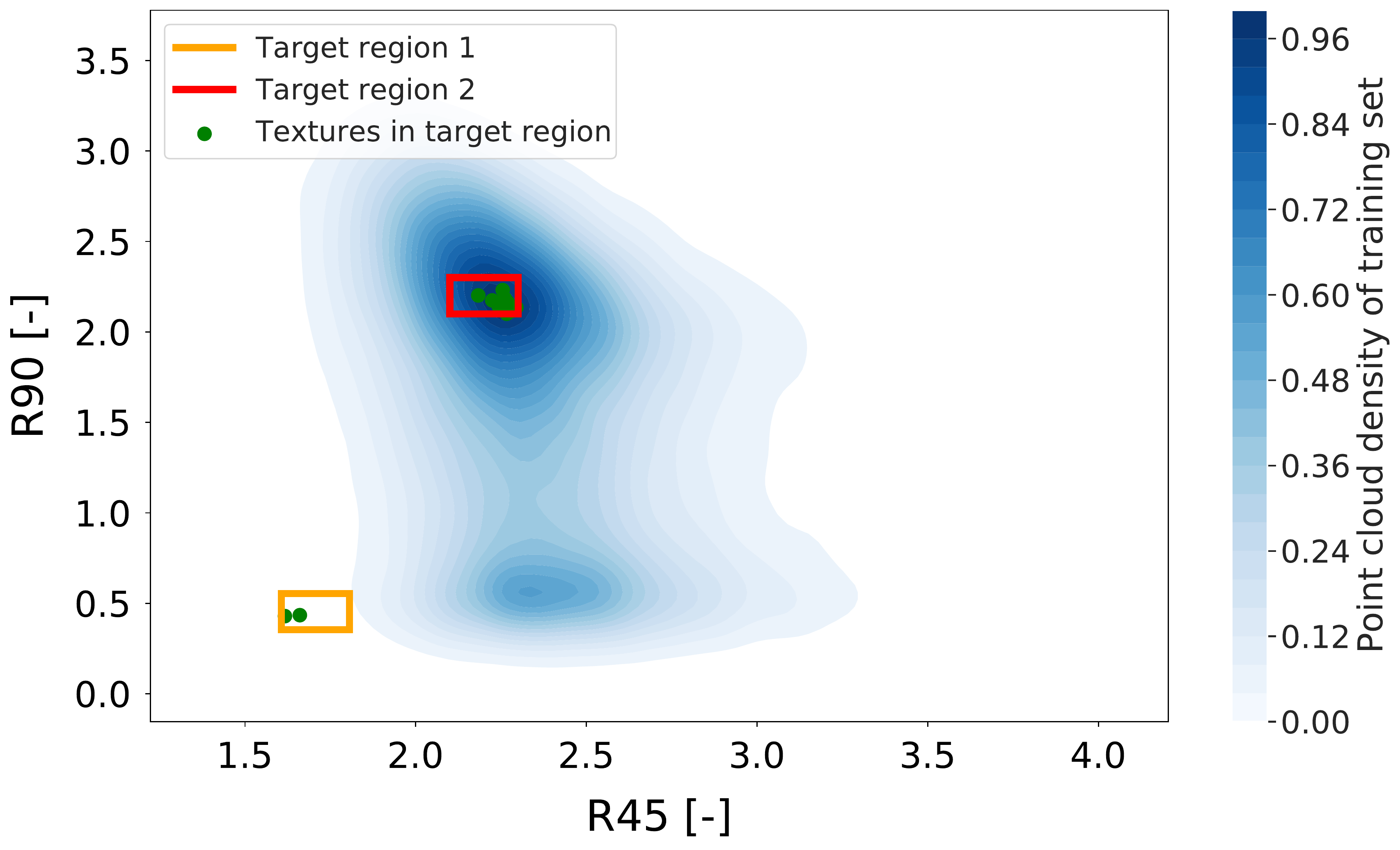}}%   
  
  \caption{Density of the training set projected on different planes of the property space: the Young's modulus at (a) 0 vs. 90 degree and (b) 45 vs. 90 degree for the R-values at (c) 0 vs. 90 degree and (d) 45 vs. 90 to rolling direction. The orange and red squares mark the projections of \textit{Target Region 1} and \textit{Target Region 2}, respectively. The green dots show the projected samples from the training set that lie inside the target region.} 
  \label{img:Target-regions-e-values}
\end{figure}

To validate the texture identification, we define two target regions in the property space, see Fig. \ref{img:Target-regions-e-values}. The first one is defined by the properties of the experimentally measured texture, which lies in a sparsely populated region and is labeled as \textit{Target Region 1}. As a consequence of its location in the sparsely populated region, the anomaly score of this texture is $0.0099$ and in the transition zone shiftet towards the generated anomalies (cf. Fig. \ref{fig:Anomalie-detection_histogram}). It is of interest if the optimizer is generally able to find a whole set of microstructures with properties in this region. The second target region represents a densely populated region located near the center of the properties point cloud and is labeled as \textit{Target Region 2}. The center of the each target region is listed in Tab. \ref{tab:Target-Regions}. The target regions are defined by adding a tolerance of $\pm 1000$ MPa to the Young's moduli and $\pm0.10$ to the R-values. As a baseline, we collect all data points from the training set, that lie inside the target regions. In \textit{Target Region 1} only two textures can be found, whereas in \textit{Target Region 2} $13$ textures can be found.

\begin{table}
  \centering
  \caption{Center points of the two target regions (TR). The Young's moduli $E$ are given in MPa, the R-values $r$ in [-].}
  \begin{tabular}{|l|c|c|c|c|c|c|}
  \hline
   TR & $E_{00}$ & $E_{45}$ & $E_{90}$ & $r_{00}$ & $r_{45}$ & $r_{90}$ \\
  \hline
  \hline
  1 & $223145$ & $207148$ & $216599$ & $1.01$ & $1.7$ & $0.4$ \\
  2 & $223000$ & $213000$ & $222000$ & $1.3$ & $2.2$ & $2.2$ \\
  \hline 
  \end{tabular}
  	\label{tab:Target-Regions}
\end{table}

To identify a diverse set of textures, we use the optimization algorithm JADE \cite{zhang2009JADE}, which is an extension of the Differential Evolution algorithm \cite{storn1997DifferentialEvolution}. Before starting the optimization via JADE, an initial population has to be selected, where 100 textures are sampled from the test set, which are approximately uniformly distributed over the property space. For the cost function, defined in Eq. \ref{eq:score-fitness}, we use the weights $\mathscr{V}_\mathrm{prop}=0.90$, $\mathscr{V}_\mathrm{valid}=0.03$ and $\mathscr{V}_\mathrm{divers}=0.07$ and scale $C_\mathrm{props}$ and $C_\mathrm{divers}$ to values between $0$ and $1$ based on the selected 100 initial a textures. The threshold $\xi_\mathrm{valid}$ is set to $0.01$ based on the maximum anomaly score in the data set, cf. Fig. \ref{fig:Anomalie-detection_histogram}. The optimization is performed for 300 iterations with a fixed population size of 100. During the optimization, all valid textures that fulfill the target properties are collected, according to the texture-property-mapping. Based on the results from the previous section, we use the trained SMTL-model with a $16$-dimensional latent feature space. The resulting textures for each target region are discussed in the following.

\subsubsection*{Target region 1} 

\begin{figure}
	\centering
  	\includegraphics[width=0.95\linewidth]{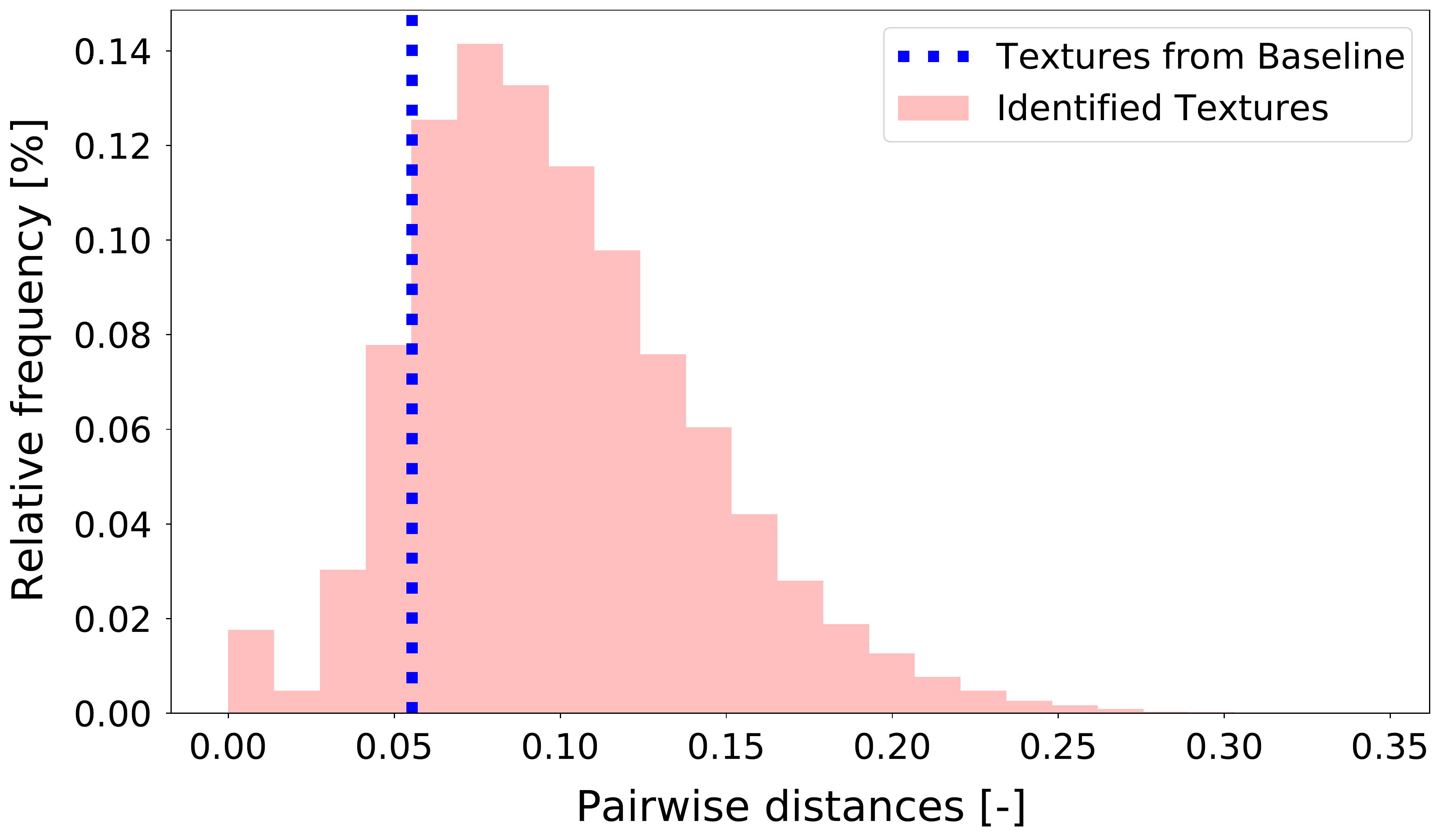}
	\caption{Histogram of pairwise SSE distances of the set of identified textures and the baseline set for \textit{Target Region 1}. The distance between the two textures from the baseline set is indicated by the dashed line.}
	\label{fig:Histogram-diversity-with-baseline_TR1}
\end{figure}

\begin{figure}
	\centering
  	\includegraphics[width=0.95\linewidth]{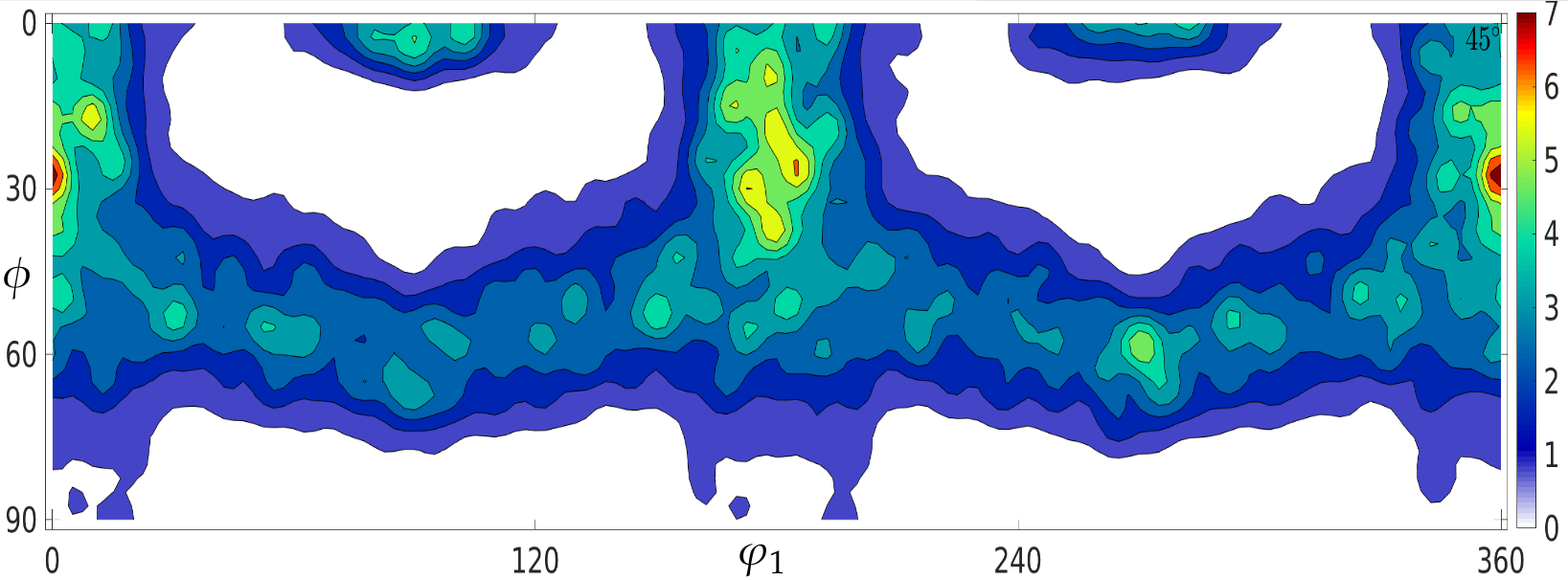}
	\caption{Texture that yields properties which are closest to the center of \textit{Target Region 1}. The plot shows the $\varphi_2=45^\circ$ section of the orientation distribution function.}
	\label{fig:Generated-microstructure-compare-plot-euler-space-phi2-45-TR1}
\end{figure}

Our approach is able to find a diverse set of textures that meet the property requirements of \textit{Target Region 1}, according to the texture-property-mapping. Fig. \ref{fig:Histogram-diversity-with-baseline_TR1} depicts the mutual distances in the latent feature space between all the found textures and between the two baseline textures. It is shown, that the set of identified textures contains 1315 diverse textures in contrary to only two in the baseline set. In order to compare the results to the experimentally measured texture, the closest texture to the center point of \textit{Target Region 1} is depicted in Fig \ref{fig:Generated-microstructure-compare-plot-euler-space-phi2-45-TR1} as a section through euler space at $\varphi_2 = 45^\circ$. By comparing the two textures, it can be seen that they are roughly the same in terms of the magnitude of the intensities and the shape of the $\alpha$- and $\gamma$-fibers. However, they also show differences in terms of smoothness and the location of the intensity peaks.

\subsubsection*{Target region 2}

\begin{figure}
	\centering
  	\includegraphics[width=0.95\linewidth]{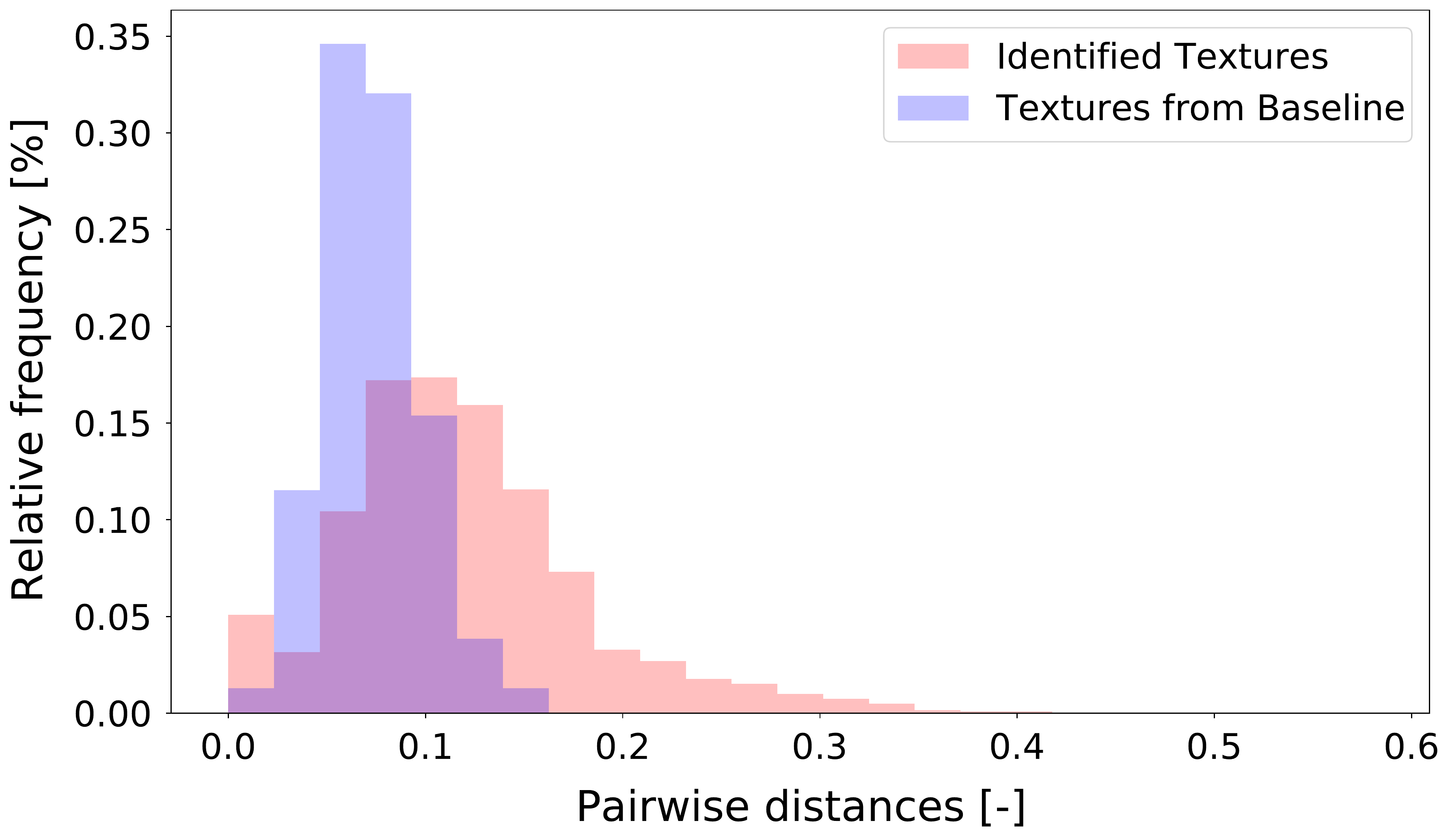}
	\caption{Histogram of mutual distances of the set of identified textures and the baseline set for \textit{Target Region 2}.}
	\label{fig:Histogram-diversity-with-baseline_TR2}
\end{figure}

\begin{figure}
	\centering
  	\subfloat[][]{\includegraphics[width=0.95\textwidth]{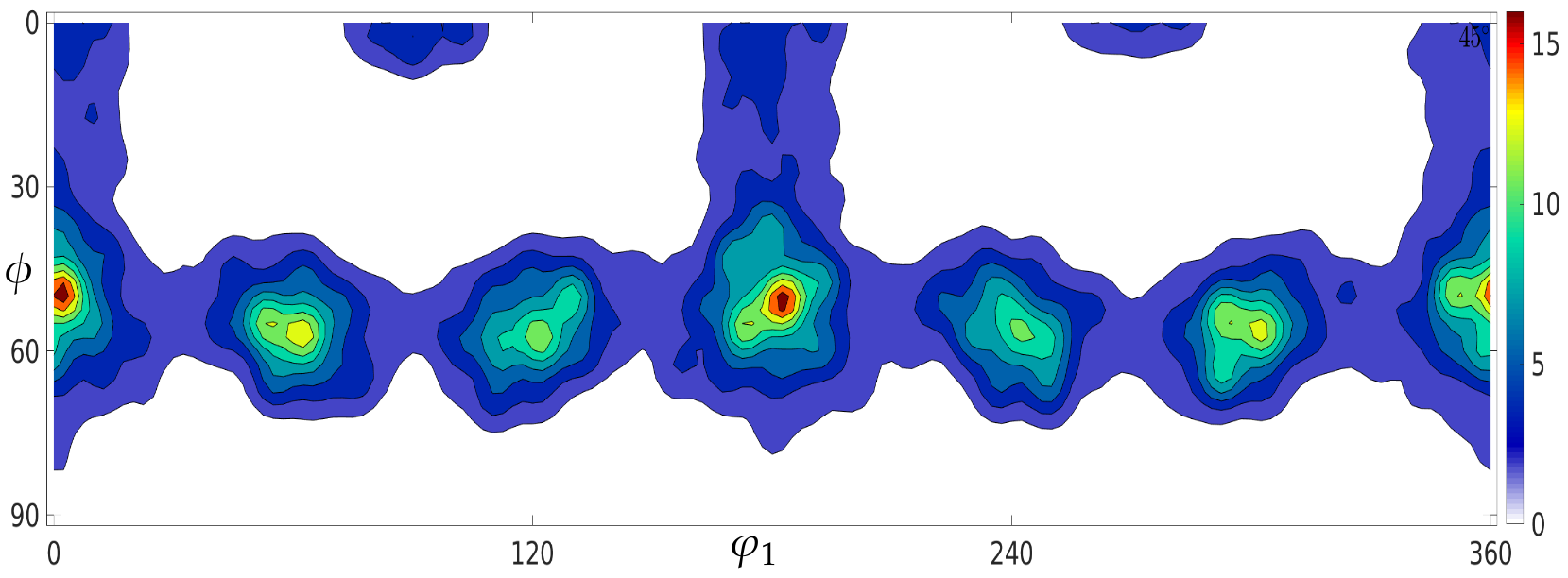}}%
  	\qquad
	\subfloat[][]{\includegraphics[width=0.95\textwidth]{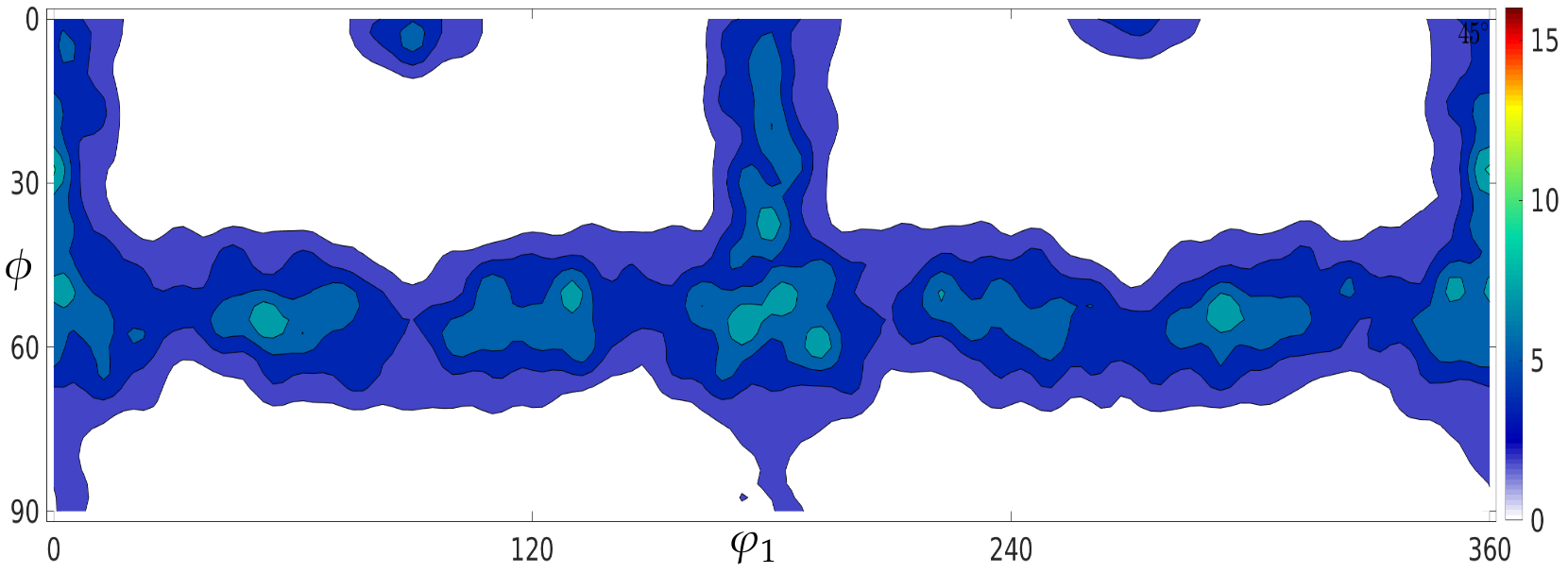}}%
	\caption{Two exemplary textures from the set of identified textures. Both plots show $\varphi_2 = 45^\circ$ sections of the respective orientation distribution functions.}
	\label{fig:Generated-microstructure-compare-plot-euler-space-phi2-45-TR2}
\end{figure}

Compared to \textit{Target Region 1}, an even more diverse set of 221 textures can be identified for \textit{Target Region 2}, which can be seen in the histogram of the mutual distances in Fig. \ref{fig:Histogram-diversity-with-baseline_TR2}. To get an idea of the differences between the textures, two exemplary textures are plotted in Fig. \ref{fig:Generated-microstructure-compare-plot-euler-space-phi2-45-TR2} as a section through the ODF in the euler space at $\varphi_2 = 45^\circ$. It can be seen that the $\alpha$- and $\gamma$-fiber of both textures differ significantly in terms of intensity. However, the locations of the intensity peaks and the thickness of the $\alpha$- and $\gamma$-fiber are similar.

\section{Discussion}
\label{sec:discussion}

% Ab methoden nur noch texture

% Ergebnisse Model
The results presented in Section \ref{sec:results_validation} show that the two tasks texture-property-mapping and validity-prediction are solved by the SMTL model. To achieve a sufficient prediction quality for both tasks in the test set as well as for the experimentally measured texture, a minimum dimensionality of the latent feature space is needed, where also the dimensionality requirements of the siamese distance preservation goal has to be considered. 16 latent features were found to be sufficient for our example task of cold rolled steel. However, the prediction error for the experimentally measured texture is higher than for the test set using the same latent feature space dimensionality. This may be explained by the fact that the corresponding property is in a texture space region with low sampling density and the model therefore is not well supported by data. This results also in an instability of the model quality depending on the dimensionality of the latent feature space in this region. This instability can be seen by studying the R-value in Tab. \ref{tab:Results-Siamese-Experimental}. By choosing the latent feature space size of $16$, also the results for the experimentally measured texture are satisfactory, especially keeping in mind that the experimentally measured texture differs naturally from the simulated data and additionally lies in a sparsely sampled region, cf. \textit{Target Region 1} in Fig. \ref{img:Target-regions-e-values}.  

% TR1
Due to the sparsity of \textit{Target Region 1}, the identification of textures in this region is challenging. Nevertheless, the optimization approach is able to identify a set of textures that contains more diverse individuals compared to the two baseline textures from the training set. Regarding the identified texture, which is closest to the experimentally measured texture in terms of properties, one can see that they are also similar in terms of crystallographic texture, what basically proofs the concept of our approach. 

The most obvious difference between both textures is smoothness. The irregular distribution of intensity peaks of the identified texture is due to the resolution of the histogram-based texture descriptor. Also the orthorhombic sample symmetry is not represented locally. However, by increasing the resolution, these two issues can be solved. Furthermore, a higher resolution of the descriptor decreases the descriptor error, which reflects the deviation between the properties of the original texture and the properties of the texture described by the descriptor. However, the choice of resolution is a trade-off between accuracy and descriptor complexity, which is why we chose an intermediate resolution to proof our approach. Generally, with the use of the SMTL model and the incorporated feature extraction, the resolution is limited only by computational power.

% TR2
Compared to \textit{Target Region 1}, the identification task for \textit{Target Region 2} seems to be less challenging as the target region is located in a densely sampled region. However, as there already exists a proper set of diverse textures in the baseline, the main challenge is to outperform the baseline set in terms of diversity. Fig. \ref{fig:Histogram-diversity-with-baseline_TR2} shows that this general materials design problem (the identification of multiple equivalent microstructures) is accomplished by the optimization approach. This is exemplarily shown when comparing two of the identified textures in Fig. \ref{fig:Generated-microstructure-compare-plot-euler-space-phi2-45-TR2} with each other: similar properties can be reached by quite different microstructures. The identification of such a highly diverse set of microstructures with similar properties is an important precondition to construct robust optimizing process control algorithms, which can then choose among multiple optimal paths leading to desired properties.
%The next step is therefore to define texture regions for processing, which can for example be done with the help of our approach, based on the generated latent features space.
% Vergleich mit kuroda ?
%In comparison to the initially mention work by Kuroda and Ikawa \cite{kuroda2004texture}, we are now able to efficiently

% The cost terms of the objective function from the optimizer have conflicting interests, because a higher diversity causes a lower validity. Nevertheless, when the textures have reached the target region, the property cost term becomes zero and the cost terms for validity and diversity are minimized.

\section{Summary and outlook}
\label{sec:conclusion}

% zusammenfassung
In this work we present an approach to solve materials design problems. The approach is based on an optimization strategy that incorporates machine learning models for mapping microstructures to properties and for assessing the validity of input microstructures in the sense of the likeliness with the underlying data. To model these tasks, we use a siamese multi-task learning neural network (SMTL). Furthermore, we incorporate feature extraction in order to transform input microstructures to a lower dimensional latent feature space where the optimizer looking for microstructures with dedicated properties can efficiently operate. 

By training the SMTL network with a dedicated loss function term, we are able to preserve the distances between microstructures in the original input space also in the latent feature space. The distance preservation allows to directly assess the diversity of the solution set found by the optimizer directly in the latent feature space and therefore enables optimizers to efficiently identify sets of diverse microstructures. By applying the approach to crystallographic texture optimization, we show the ability to identify diverse sets of textures that lie within given bounds of properties. Such sets of textures form the input of optimal processing control approaches like in \cite{dornheim2020structureguided}.

% future work
In the present work, we applied our approach on data from mean-field simulations. The next step is to apply the approach on spatially resolved data from full-field simulations. The proposed methods can be easily extended for this task by modifying the encoder part of the SMTL network. However, the problem arises that typically fewer data can be generated via full-field simulations. Nevertheless, such sparse high quality data can be used to support the modeling with lower quality data. Concepts to incoorporate such multi-fidelity data fusion approaches \cite{batra2021accurate} in our SMTL model learning will be considered in the future. 

%Further investigations on machine learning methods for the identification of optimal textures will also be subject of future work. Methods that are worth considering are generative approaches such as conditional variational autoencoders \cite{Kingma2014-CVAE} or conditional generative adversarial nets \cite{mirza2014conditionalGAN}.%, where the desired properties represent the condition-dependent input. 

\section*{Data availability}

The data used to validate the SMTL approach is made available via the repository Fordatis at \url{https://fordatis.fraunhofer.de/handle/fordatis/204} \cite{data2021}.

\section*{Author Contributions} TI developed and implemented the SMTL-framework as well as the optimization concept for materials design. LM set up the rolling texture generation model, implemented the properties calculation framework on the basis of the Taylor model and generated the used texture-property data. JD contributed theory and implementation of the crystallographic texture descriptor and distance function. NL supported in terms of machine learning. DH supported in terms of materials science with focus on the modelling of polycristalline material behavior. All authors formulated the scientific challenge, contributed to the solution strategy and contributed to the discussion and the summary and outlook part of the paper.
\newline

\acknowledgements{
%\section*{\ackname}
%\textbf{Acknowledgements}
The authors would like to thank the German Research Foundation (DFG) for funding the presented work, which was carried out within the research project number 415804944: ’Taylored Material Properties via Microstructure Optimization: Machine Learning for Modelling and Inversion of Structure-Property-Relationships and the Application to Sheet Metals’. Also, we would like to thank Jan Pagenkopf for providing the crystal plasticity routine on which the implemented Taylor-type material model is based.}

%\begin{acknowledgements}
%The authors would like to thank the German Research Foundation (DFG) for funding the presented work, which was carried out within the research project number 415804944: %’Taylored Material Properties via Microstructure Optimization: Machine Learning for Modelling and Inversion of Structure-Property-Relationships and the Application to %Sheet Metals’. Also, we would like to thank Jan Pagenkopf for providing the crystal plasticity routine on which the implemented Taylor-type material model is based.
%\end{acknowledgements}

\bibliographystyle{ieeetr}
\bibliography{paper}   % name your BibTeX data base

\begin{thebibliography}{10}

\bibitem{olson1997computational}
G.~B. Olson, ``Computational design of hierarchically structured materials,''
  {\em Science}, vol.~277, no.~5330, pp.~1237--1242, 1997.

\bibitem{panchal2013key}
J.~H. Panchal, S.~R. Kalidindi, and D.~L. McDowell, ``Key computational
  modeling issues in integrated computational materials engineering,'' {\em
  Computer-Aided Design}, vol.~45, no.~1, pp.~4--25, 2013.

\bibitem{ramprasad2017machine}
R.~Ramprasad, R.~Batra, G.~Pilania, A.~Mannodi-Kanakkithodi, and C.~Kim,
  ``Machine learning in materials informatics: recent applications and
  prospects,'' {\em npj Computational Materials}, vol.~3, no.~1, pp.~1--13,
  2017.

\bibitem{dornheim2020structureguided}
J.~Dornheim, L.~Morand, S.~Zeitvogel, T.~Iraki, N.~Link, and D.~Helm, ``Deep
  reinforcement learning methods for structure-guided processing path
  optimization,'' {\em Journal of Intelligent Manufacturing}, 2021.

\bibitem{mcdowell2007simulation}
D.~L. McDowell, ``Simulation-assisted materials design for the concurrent
  design of materials and products,'' {\em JOM}, vol.~59, no.~9, pp.~21--25,
  2007.

\bibitem{adams2001microstructure}
B.~L. Adams, A.~Henrie, B.~Henrie, M.~Lyon, S.~Kalidindi, and H.~Garmestani,
  ``Microstructure-sensitive design of a compliant beam,'' {\em Journal of the
  Mechanics and Physics of Solids}, vol.~49, no.~8, pp.~1639--1663, 2001.

\bibitem{fullwood2010microstructure}
D.~T. Fullwood, S.~R. Niezgoda, B.~L. Adams, and S.~R. Kalidindi,
  ``Microstructure sensitive design for performance optimization,'' {\em
  Progress in Materials Science}, vol.~55, no.~6, pp.~477--562, 2010.

\bibitem{kalidindi2004microstructure}
S.~R. Kalidindi, J.~R. Houskamp, M.~Lyons, and B.~L. Adams, ``Microstructure
  sensitive design of an orthotropic plate subjected to tensile load,'' {\em
  International Journal of Plasticity}, vol.~20, no.~8-9, pp.~1561--1575, 2004.

\bibitem{bunge2013texture}
H.-J. Bunge, {\em Texture Analysis in Materials Science: Mathematical Methods}.
\newblock Burlington: {Elsevier Science}, digitale~ed., 2013.

\bibitem{paulson2017reduced}
N.~H. Paulson, M.~W. Priddy, D.~L. McDowell, and S.~R. Kalidindi,
  ``Reduced-order structure-property linkages for polycrystalline
  microstructures based on 2-point statistics,'' {\em Acta Materialia},
  vol.~129, pp.~428--438, 2017.

\bibitem{gupta2015structure}
A.~Gupta, A.~Cecen, S.~Goyal, A.~K. Singh, and S.~R. Kalidindi,
  ``Structure--property linkages using a data science approach: application to
  a non-metallic inclusion/steel composite system,'' {\em Acta Materialia},
  vol.~91, pp.~239--254, 2015.

\bibitem{JUNG2019-MDS}
J.~Jung, J.~I. Yoon, H.~K. Park, J.~Y. Kim, and H.~S. Kim, ``An efficient
  machine learning approach to establish structure-property linkages,'' {\em
  Computational Materials Science}, vol.~156, pp.~17--25, 2019.

\bibitem{van2009dimensionality}
L.~Van Der~Maaten, E.~Postma, J.~Van~den Herik, {\em et~al.}, ``Dimensionality
  reduction: a comparative,'' {\em J Mach Learn Res}, vol.~10, no.~66-71,
  p.~13, 2009.

\bibitem{liu2015predictive}
R.~Liu, A.~Kumar, Z.~Chen, A.~Agrawal, V.~Sundararaghavan, and A.~Choudhary,
  ``A predictive machine learning approach for microstructure optimization and
  materials design,'' {\em Scientific Reports}, vol.~5, no.~1, pp.~1--12, 2015.

\bibitem{paul2019microstructure}
A.~Paul, P.~Acar, W.-k. Liao, A.~Choudhary, V.~Sundararaghavan, and A.~Agrawal,
  ``Microstructure optimization with constrained design objectives using
  machine learning-based feedback-aware data-generation,'' {\em Computational
  Materials Science}, vol.~160, pp.~334--351, 2019.

\bibitem{kuroda2004texture}
M.~Kuroda and S.~Ikawa, ``Texture optimization of rolled aluminum alloy sheets
  using a genetic algorithm,'' {\em Materials Science and Engineering: A},
  vol.~385, no.~1-2, pp.~235--244, 2004.

\bibitem{goldberg1990real}
D.~Goldberg, ``Real-coded genetic algorithms, virtual alphabets and blocking,''
  {\em Complex Systems}, vol.~5, 1991.

\bibitem{herrera1998tackling}
F.~Herrera, M.~Lozano, and J.~L. Verdegay, ``Tackling real-coded genetic
  algorithms: Operators and tools for behavioural analysis,'' {\em Artificial
  Intelligence Review}, vol.~12, no.~4, pp.~265--319, 1998.

\bibitem{simpson2001metamodels}
T.~W. Simpson, J.~Poplinski, P.~N. Koch, and J.~K. Allen, ``Metamodels for
  computer-based engineering design: survey and recommendations,'' {\em
  Engineering with computers}, vol.~17, no.~2, pp.~129--150, 2001.

\bibitem{Jung2019-ocsvm}
J.~Jung, J.~I. Yoon, S.-J. Park, J.-Y. Kang, G.~L. Kim, Y.~H. Song, S.~T. Park,
  K.~W. Oh, and H.~S. Kim, ``Modelling feasibility constraints for materials
  design: Application to inverse crystallographic texture problem,'' {\em
  Computational Materials Science}, vol.~156, pp.~361--367, 2019.

\bibitem{Chandola2009-AnomalyDetection}
V.~Chandola, A.~Banerjee, and V.~Kumar, ``Anomaly detection: A survey,'' {\em
  ACM Computing Surveys (CSUR)}, vol.~41, no.~3, 2009.

\bibitem{Deep-Learning-for-Anomaly-Detection-Survey-2019}
R.~Chalapathy and S.~Chawla, ``Deep learning for anomaly detection: {A}
  survey,'' {\em arXiv:1901.03407}, 2019.

\bibitem{HintonSalakhutdinov2006b}
G.~E. Hinton and R.~R. Salakhutdinov, ``Reducing the dimensionality of data
  with neural networks,'' {\em Science}, vol.~313, no.~5786, pp.~504--507,
  2006.

\bibitem{Anomaly-Detection-Using-Autoencoders-Sakurada-2014}
M.~Sakurada and T.~Yairi, ``Anomaly detection using autoencoders with nonlinear
  dimensionality reduction,'' in {\em Proceedings of the MLSDA 2014 2nd
  Workshop on Machine Learning for Sensory Data Analysis}, pp.~4--11, 2014.

\bibitem{caruana1997mtl}
R.~Caruana, ``Multitask learning,'' {\em Machine Learning}, vol.~28, no.~1,
  p.~41–75, 1997.

\bibitem{bromley1993siamese}
J.~Bromley, I.~Guyon, Y.~LeCun, E.~S\"{a}ckinger, and R.~Shah, ``Signature
  verification using a "{S}iamese" time delay neural network,'' in {\em
  Advances in Neural Information Processing Systems 6}, pp.~737--744, 1993.

\bibitem{Krizhevsky2012-CNN}
A.~Krizhevsky, I.~Sutskever, and G.~E. Hinton, ``Imagenet classification with
  deep convolutional neural networks,'' in {\em Advances in Neural Information
  Processing Systems 25}, pp.~1106--1114, 2012.

\bibitem{cecen2018material}
A.~Cecen, H.~Dai, Y.~C. Yabansu, S.~R. Kalidindi, and L.~Song, ``Material
  structure-property linkages using three-dimensional convolutional neural
  networks,'' {\em Acta Materialia}, vol.~146, pp.~76--84, 2018.

\bibitem{Krogh1991Weight-Decay}
A.~Krogh and J.~A. Hertz, ``A simple weight decay can improve generalization,''
  in {\em Advances in Neural Information Processing Systems 4}, p.~950–957,
  1991.

\bibitem{Hinton1987Weight-Decay}
G.~E. Hinton, ``Learning translation invariant recognition in a massively
  parallel networks,'' in {\em International Conference on Parallel
  Architectures and Languages Europe}, pp.~1--13, Springer, 1987.

\bibitem{Chicco2021}
D.~Chicco, ``Siamese neural networks: An overview,'' {\em Artificial Neural
  Networks}, pp.~73--94, 2021.

\bibitem{A-Siamese-Autoencoder-Preserving-Distances-2017}
L.~V. Utkin, V.~S. Zaborovsky, A.~A. Lukashin, S.~G. Popov, and A.~V.
  Podolskaja, ``A siamese autoencoder preserving distances for anomaly
  detection in multi-robot systems,'' in {\em 2017 International Conference on
  Control, Artificial Intelligence, Robotics \& Optimization (ICCAIRO)},
  pp.~39--44, IEEE, 2017.

\bibitem{Kruskal1964MultidimensionalSB}
J.~Kruskal, ``Multidimensional scaling by optimizing goodness of fit to a
  nonmetric hypothesis,'' {\em Psychometrika}, vol.~29, pp.~1--27, 1964.

\bibitem{Multidimensional-Scaling-Cox2008}
M.~A. Cox and T.~F. Cox, ``Multidimensional scaling,'' in {\em Handbook of data
  visualization}, pp.~315--347, Berlin Heidelberg: Springer, 2008.

\bibitem{hansen1978tables}
J.~Hansen, J.~Pospiech, and K.~L{\"u}cke, {\em Tables for texture analysis of
  cubic crystals}.
\newblock Berlin Heidelberg New York: Springer, 1978.

\bibitem{Huynh2009}
D.~Q. Huynh, ``{Metrics for 3D rotations: Comparison and analysis},'' {\em
  Journal of Mathematical Imaging and Vision}, vol.~35, no.~2, pp.~155--164,
  2009.

\bibitem{Pele2010chi-square-distance}
O.~Pele and M.~Werman, ``The quadratic-chi histogram distance family,'' in {\em
  European Conference on Computer Vision}, pp.~749--762, Springer, 2010.

\bibitem{Quey2018}
R.~Quey, A.~Villani, and C.~Maurice, ``Nearly uniform sampling of crystal
  orientations,'' {\em Journal of Applied Crystallography}, vol.~51, no.~4,
  pp.~1162--1173, 2018.

\bibitem{quey2011large}
R.~Quey, P.~Dawson, and F.~Barbe, ``Large-scale 3d random polycrystals for the
  finite element method: Generation, meshing and remeshing,'' {\em Computer
  Methods in Applied Mechanics and Engineering}, vol.~200, no.~17-20,
  pp.~1729--1745, 2011.

\bibitem{ray1994cold}
R.~Ray, J.~J. Jonas, and R.~Hook, ``Cold rolling and annealing textures in low
  carbon and extra low carbon steels,'' {\em International Materials Reviews},
  vol.~39, no.~4, pp.~129--172, 1994.

\bibitem{kocks1998texture}
U.~F. Kocks, C.~N. Tom{\'e}, and H.-R. Wenk, {\em Texture and anisotropy:
  preferred orientations in polycrystals and their effect on materials
  properties}.
\newblock Cambridge New York Melbourne: Cambridge University Press, 1998.

\bibitem{von1986investigation}
U.~Von~Schlippenbach, F.~Emren, and K.~L{\"u}cke, ``Investigation of the
  development of the cold rolling texture in deep drawing steels by
  odf-analysis,'' {\em Acta metallurgica}, vol.~34, no.~7, pp.~1289--1301,
  1986.

\bibitem{delannay1999new}
L.~Delannay, P.~Van~Houtte, and A.~Van~Bael, ``New parameter model for texture
  description in steel sheets,'' {\em Texture, Stress, and Microstructure},
  vol.~31, no.~3, pp.~151--175, 1999.

\bibitem{Kalidindi.1992}
S.~R. Kalidindi, C.~A. Bronkhorst, and L.~Anand, ``Crystallographic texture
  evolution in bulk deformation processing of fcc metals,'' {\em Journal of the
  Mechanics and Physics of Solids}, vol.~40, no.~3, pp.~537--569, 1992.

\bibitem{eghtesad2020high}
A.~Eghtesad and M.~Knezevic, ``High-performance full-field crystal plasticity
  with dislocation-based hardening and slip system back-stress laws:
  Application to modeling deformation of dual-phase steels,'' {\em Journal of
  the Mechanics and Physics of Solids}, vol.~134, p.~103750, 2020.

\bibitem{Rice.1971}
J.~R. Rice, ``Inelastic constitutive relations for solids: {A}n
  internal-variable theory and its application to metal plasticity,'' {\em
  Journal of the Mechanics and Physics of Solids}, vol.~19, no.~6,
  pp.~433--455, 1971.

\bibitem{Asaro.1985}
R.~J. Asaro and A.~Needleman, ``Overview no. 42 texture development and strain
  hardening in rate dependent polycrystals,'' {\em Acta Metallurgica}, vol.~33,
  no.~6, pp.~923--953, 1985.

\bibitem{Pagenkopf.2016}
J.~Pagenkopf, A.~Butz, M.~Wenk, and D.~Helm, ``Virtual testing of dual-phase
  steels: Effect of martensite morphology on plastic flow behavior,'' {\em
  Materials Science and Engineering: A}, vol.~674, pp.~672--686, 2016.

\bibitem{baiker2014determination}
M.~Baiker, D.~Helm, and A.~Butz, ``Determination of mechanical properties of
  polycrystals by using crystal plasticity and numerical homogenization
  schemes,'' {\em Steel Research International}, vol.~85, no.~6, pp.~988--998,
  2014.

\bibitem{Tome.1984}
C.~Tome, G.~R. Canova, U.~F. Kocks, N.~Christodoulou, and J.~J. Jonas, ``The
  relation between macroscopic and microscopic strain hardening in f.c.c.
  polycrystals,'' {\em Acta Metallurgica}, vol.~32, no.~10, pp.~1637--1653,
  1984.

\bibitem{Pagenkopf.2019}
J.~Pagenkopf, {\em Bestimmung der {P}lastischen {A}nisotropie von
  {B}lechwerkstoffen durch ortsaufgel{\"o}ste {S}imulationen auf
  {G}ef{\"u}geebene}.
\newblock PhD thesis, {Fakult{\"a}t f{\"u}r Maschinenbau des Karlsruher
  Instituts f{\"u}r Technologie (KIT)}, 2019.

\bibitem{McKay.1979}
M.~D. McKay, R.~J. Beckman, and W.~J. Conover, ``A comparison of three methods
  for selecting values of input variables in the analysis of output from a
  computer code,'' {\em Technometrics}, vol.~21, no.~2, p.~239, 1979.

\bibitem{lucke1970rolling}
M.~H{\"o}lscher, D.~Raabe, and K.~L{\"u}cke, ``Rolling and recrystallization
  textures of bcc steels,'' {\em Steel Research}, vol.~62, no.~12,
  pp.~567--575, 1991.

\bibitem{inagaki1972development}
H.~Inagaki and T.~Suda, ``The development of rolling textures in low-carbon
  steels,'' {\em Texture, Stress, and Microstructure}, vol.~1, no.~2,
  pp.~129--140, 1972.

\bibitem{klinkenberg1992effects}
C.~Klinkenberg, D.~Raabe, and K.~L{\"u}cke, ``Influence of volume fraction and
  dispersion rate of grain-boundary cementite on the cold-rolling textures of
  low-carbon steel,'' {\em Steel Research}, vol.~63, no.~6, pp.~263--269, 1992.

\bibitem{kestens2016texture}
L.~Kestens and H.~Pirgazi, ``Texture formation in metal alloys with cubic
  crystal structures,'' {\em Materials Science and Technology}, vol.~32,
  no.~13, pp.~1303--1315, 2016.

\bibitem{das2017calculation}
A.~Das, ``Calculation of crystallographic texture of bcc steels during cold
  rolling,'' {\em Journal of Materials Engineering and Performance}, vol.~26,
  no.~6, pp.~2708--2720, 2017.

\bibitem{Schreijag.2012}
S.~Schreij{\"a}g, {\em Microstructure and Mechanical Behavior of Deep Drawing
  {DC04} Steel at Different Length Scales}.
\newblock PhD thesis, {Fakult{\"a}t f{\"u}r Maschinenbau des Karlsruher
  Instituts f{\"u}r Technologie (KIT)}, 2012.

\bibitem{bachmann2010}
F.~Bachmann, R.~Hielscher, and H.~Schaeben, ``Texture analysis with mtex –
  free and open source software toolbox,'' in {\em Solid State Phenomena},
  vol.~160, pp.~63--68, 3 2010.

\bibitem{tensorflow2015-whitepaper}
M.~Abadi, A.~Agarwal, P.~Barham, E.~Brevdo, Z.~Chen, C.~Citro, G.~S. Corrado,
  A.~Davis, J.~Dean, M.~Devin, S.~Ghemawat, I.~Goodfellow, A.~Harp, G.~Irving,
  M.~Isard, Y.~Jia, R.~Jozefowicz, L.~Kaiser, M.~Kudlur, J.~Levenberg,
  D.~Man\'{e}, R.~Monga, S.~Moore, D.~Murray, C.~Olah, M.~Schuster, J.~Shlens,
  B.~Steiner, I.~Sutskever, K.~Talwar, P.~Tucker, V.~Vanhoucke, V.~Vasudevan,
  F.~Vi\'{e}gas, O.~Vinyals, P.~Warden, M.~Wattenberg, M.~Wicke, Y.~Yu, and
  X.~Zheng, ``{TensorFlow}: Large-scale machine learning on heterogeneous
  systems,'' 2015.
\newblock white paper.

\bibitem{Glorot2010WeightInit}
X.~Glorot and Y.~Bengio, ``Understanding the difficulty of training deep
  feedforward neural networks,'' in {\em Proceedings of the 13th International
  Conference on Artificial Intelligence and Statistics}, pp.~249--256, JMLR
  Workshop and Conference Proceedings, 2010.

\bibitem{Bergstra2012GridSearch}
J.~Bergstra and Y.~Bengio, ``Random search for hyper-parameter optimization,''
  {\em Journal of Machine Learning Research}, vol.~13, no.~10, pp.~281--305,
  2012.

\bibitem{Prechelt2012EarlyStopping}
L.~Prechelt, ``Early stopping-but when?,'' in {\em Neural Networks: Tricks of
  the trade}, pp.~55--69, Springer, 1998.

\bibitem{Kingma2015Adam}
D.~P. Kingma and J.~Ba, ``Adam: {A} method for stochastic optimization,'' in
  {\em 3rd International Conference on Learning Representations}, 2015.

\bibitem{zhang2009JADE}
J.~{Zhang} and A.~C. {Sanderson}, ``Jade: Adaptive differential evolution with
  optional external archive,'' {\em IEEE Transactions on Evolutionary
  Computation}, vol.~13, no.~5, pp.~945--958, 2009.

\bibitem{storn1997DifferentialEvolution}
R.~Storn and K.~Price, ``Differential evolution – a simple and efficient
  heuristic for global optimization over continuous spaces,'' {\em Journal of
  Global Optimization}, vol.~11, no.~4, p.~341–359, 1997.

\bibitem{batra2021accurate}
R.~Batra, ``Accurate machine learning in materials science facilitated by using
  diverse data sources,'' {\em Nature}, vol.~589, 2021.

\bibitem{data2021}
L.~Morand, T.~Iraki, J.~Dornheim, J.~Pagenkopf, and D.~Helm, ``Artificially
  generated crystallographic textures of steel sheets and their corresponding
  properties calculated by a {T}aylor-type crystal plasticity model,'' 2021.
\newblock Data sets available at
  \url{https://fordatis.fraunhofer.de/handle/fordatis/204}.

\end{thebibliography}

\end{document}